\documentclass[12pt]{spieman}  
\usepackage{amsmath,amsfonts,amssymb}
\usepackage{graphicx}
\usepackage{tocloft}
\usepackage{url}
\usepackage{changes}
\definechangesauthor[name={Andrew Benson},color=red]{AJB}
\definechangesauthor[name={Matt Bradford},color=magenta]{CMB}

\newcommand{\asec}{$^{\prime\prime}$}

\newcommand{\nepunits}{W\,Hz$^{-1/2}$}
\newcommand{\sxxunits}{\,Hz$^{-1}$}
\newcommand{\lt}{<}
\newcommand{\gt}{>}

\newcommand{\mm}{$\,\rm\mu m$}

\newcommand{\micron}{$\mu$m}

\title{Galaxy Evolution Probe}

\author[a*]{Jason Glenn}
\author[b]{Charles M. Bradford}
\author[c]{Erik Rosolowsky}
\author[b]{Rashied Amini}
\author[d]{Katherine Alatalo}
\author[e]{Lee Armus}
\author[f]{Andrew J. Benson}
\author[b]{Tzu-Ching Chang}
\author[g]{Jeremy Darling}
\author[b]{Peter K. Day}
\author[h]{Jeanette Domber}
\author[i]{Duncan Farrah}
\author[j]{Brandon Hensley}
\author[h]{Sarah Lipscy}
\author[b]{Bradley Moore}
\author[k]{Seb Oliver}
\author[g]{Joanna Perido}
\author[b]{David Redding}
\author[b]{Michael Rodgers}
\author[k]{Raphael Shirley}
\author[l]{Howard A. Smith}
\author[b]{John B. Steeves}
\author[m]{Carole Tucker}
\author[n]{Jonas Zmuidzinas}

\affil[a]{NASA Goddard Space Flight Center, Code 665,  Greenbelt, MD USA}
\affil[b]{Jet Propulsion Lab, Pasadena, CA USA}
\affil[c]{2-115 Centennial Ctr For Interdisciplinary SCS II, University of Alberta, Canada}
\affil[d]{Space Telescope Science Institute, MD USA}
\affil[e]{Infrared Processing and Analysis Center, California Institute of Technology, MC 314-6, Pasadena CA}
\affil[f]{Carnegie Observatories,  Pasadena, CA USA}
\affil[g]{Center for Astrophysics and Space Astronomy, University of Colorado, 389-UCB, Boulder, CO USA}
\affil[h]{Ball Aerospace, CO USA}
\affil[i]{University of Hawaii Manoa, Hawaii USA}
\affil[j]{Department of Astrophysical Sciences, Princeton University, NJ USA}
\affil[k]{Astronomy Centre, University of Sussex, United Kingdom}
\affil[l]{Center for Astrophysics | Harvard and Smithsonian, MS-65, Cambridge, MA USA}
\affil[m]{School of Physics and Astronomy, Cardiff University, United Kingdom}
\affil[n]{The Division of Physics, Mathematics, and Astronomy, California Institute of Technology, Pasadena, CA USA}

\cftpagenumbersoff{figure}
\cftpagenumbersoff{table} 
\begin{document} 
\maketitle

\begin{abstract}
The Galaxy Evolution Probe (GEP) is a concept for a mid- and far-infrared space observatory to measure key properties of large samples of galaxies with large and unbiased surveys.  
GEP will attempt to achieve zodiacal light and Galactic dust emission photon background-limited observations by utilizing a 6 Kelvin, 2.0 meter primary mirror and sensitive arrays of kinetic inductance detectors.  It will have two instrument modules:  a 10 – 400 $\mu$m hyperspectral imager with spectral resolution $R = \lambda/\Delta\lambda \ge 8$ (GEP-I) and a 24 – 193 $\mu$m, $R = 200$ grating spectrometer (GEP-S).  GEP-I surveys will identify star-forming galaxies via their thermal dust emission and simultaneously measure redshifts using polycyclic aromatic hydrocarbon emission lines.  Galaxy luminosities derived from star formation and nuclear supermassive black hole accretion will be measured for each source, enabling the cosmic star formation history to be measured to much greater precision than previously possible.  Using optically thin far-infrared fine-structure lines, surveys with GEP-S will measure the growth of metallicity in the hearts of galaxies over cosmic time and extraplanar gas will be mapped in spiral galaxies in the local universe to investigate feedback processes.  The science case and mission architecture designed to meet the science requirements are described, and the kinetic inductance detector and readout electronics state of the art and needed developments are described.
This paper supersedes the GEP concept study report cited in it by providing new content, including:  a summary of recent mid-infrared KID development, a discussion of microlens array fabrication for mid-infrared KIDs, and additional context for galaxy surveys.  The reader interested in more technical details may want to consult the concept study report.
\end{abstract}


{\noindent \footnotesize\textbf{*}Corresponding author,  \linkable{jason.glenn@nasa.gov} }

\begin{spacing}{2}   

\begin{center}
\singlespacing{
    Appears in {\it Glenn, J., et al.  Galaxy Evolution Probe, J. Astron. Telesc. Instrum. Syst. 7(3), 034004 (2021), doi: 10.1117/1.JATIS.7.3.034004. }
    }
\end{center}

\section{Introduction}
\label{sect:intro}  

\subsection{The Galaxy Evolution Probe in Context} 

Tracing the mass assembly history of galaxies is an essential component of understanding the origins of the Hubble sequence and of using galaxies as cosmological probes of dark matter and dark energy. Observational and theoretical work over the past three decades have established a baseline framework for galaxy assembly in a cosmological context. These studies have shown that stellar and black hole mass assembly varies substantially with redshift, increasing by more than an order of magnitude between the local universe and $z=2$ \cite{shank09,madau2014cosmic,speagle14}, though the behavior at higher redshifts is uncertain \cite{tasca15}. Environment and large-scale structure profoundly affect galaxy assembly, with factors such as local galaxy density and dark matter halo properties known to play key roles in shaping the galaxy mass function \cite{dekel2006galaxy,blanton09,behroo13,schaye15}. Finally, there is a complex and subtle relationship between star formation and active galactic nucleus (AGN) activity \cite{fabian12feedback}, including both positive and negative feedback effects \cite{croton2006many,farrah12,silk2013unleashing,cicone14}. Infrared surveys from facilities including {\it IRAS} \cite{neugebauer1984infrared}, 
{\it ISO} \cite{kessler1996infrared}, {\it AKARI} \cite{murakami2007infrared}, 
{\it Herschel }\cite{pilb10}, {\it WISE} \cite{wright10}, 
and {\it Spitzer} \cite{wer04}, have played fundamental roles in these studies, since a substantial fraction of all galaxies over the history of the universe were permeated with dust through their most active evolutionary stages. 

Building upon decades of observational and theoretical work, collectively a data-driven, self-consistent model for galaxy evolution that starts from cosmology, incorporates stellar dynamics and evolution, and includes interstellar processes, star formation, and supermassive black hole growth is within reach. 
Achieving this goal requires addressing the questions that have arisen from previous and current generations of multiwavelength surveys.  These include:
What was the role of feedback from black holes and stars themselves in regulating star-formation? How did galaxies' external environments and internal contents influence their evolutionary trajectories? When were the Universe’s heavy elements formed by stars in galaxies and how did they escape into the circumgalactic and intergalactic medium? 

Answering these questions will require large panchromatic surveys measuring bulk properties of hundreds of thousands to millions of galaxies over most of cosmic history paired with detailed high angular and spectral resolution studies of gas and star formation in individual galaxies.  Such large samples are needed to precisely disentangle the effects of redshift and environment in driving galaxy assembly.  High-resolution observations of representative and outlier galaxies identified in surveys trace detailed physical processes on a galaxy-by-galaxy basis, enabling astrophysical processes to be mapped onto cosmological processes.


New mid- and far-infrared observations enabled by the rapid advances in infrared detectors and technology \cite{farrah2019far} are a vital component of these next-generation surveys. 
They require sensitivity to detect Milky Way-type galaxies at $z = 2$ ($L^* \sim 10^{12}$ L$_{\odot}$), prior to when most stellar mass had been assembled. 
Star formation rates and supermassive black hole accretion rates should be measured over a full range of cosmic environments, from isolated field galaxies, to galaxies in groups and in massive clusters. Measuring the heating of dust and gas by star formation and active galactic nuclei (AGN) is needed to understand how stellar and supermassive black hole growth were linked over cosmic time. Spectral mapping capability is needed for unbiased spatial-spectral surveys,  line luminosities and intensity mapping, and line mapping of nearby galaxies to measure gas column densities, ionization parameters, and metallicities with tracers unaffected by dust obscuration. 
Crucially,  galaxies detected by their dust continuum emission must have measured redshifts so that their epochs and luminosity distances are known and so that their multiwavelength counterparts can be identified. 

The Galaxy Evolution Probe (GEP) is a NASA Astrophysics Probe concept that capitalizes on new detector capability to address these questions with a powerful mid- and far-infrared toolset. 
GEP will 
measure star-formation rates and detect AGN even under conditions of heavy dust extinction. It will measure supermassive black hole accretion rates to address the connection between the masses of stellar populations and supermassive black holes. The same observations will measure metallicities with extinction-free tracers to observe growth of metals over the last 2/3 of the Universe’s age. In nearby galaxies, GEP will observe feedback between star-formation, AGN, and the interstellar medium to understand the processes that regulate star-formation. Mapping nearby galaxies and the Galactic interstellar medium will reveal the energy balance by measuring the total interstellar material mass, ionization state, and the local radiation field using fine-structure transitions of ions, polycyclic aromatic hydrocarbon (PAH) molecules, and the mid-infrared dust continuum.

\subsection{The GEP Concept}

\begin{figure}
\begin{center}
\begin{tabular}{c}
\includegraphics[height=5.5cm]{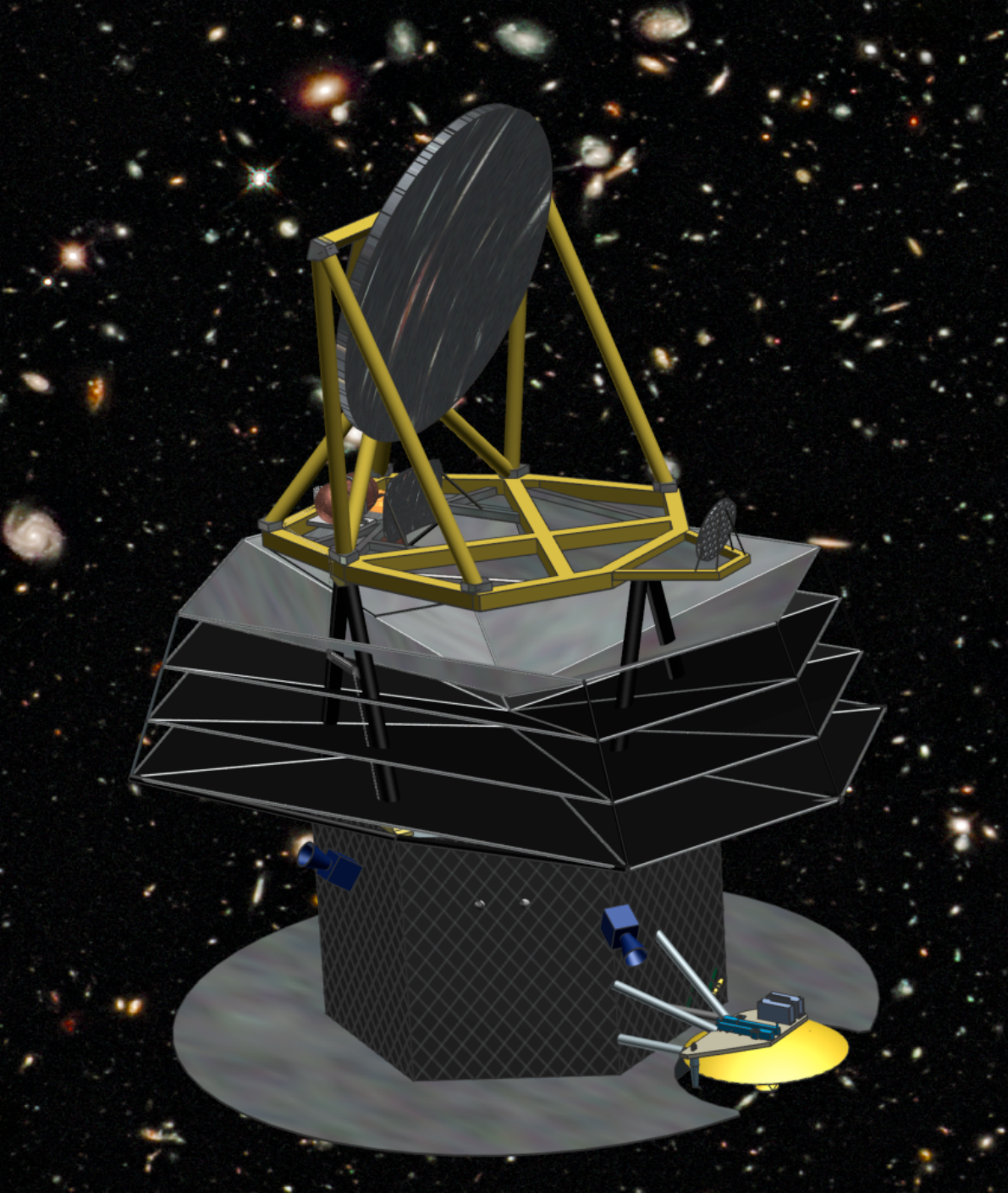}
\end{tabular}
\end{center}
\caption 
{ \label{fig:GEPRendering}
Engineering rendering of GEP. From top to bottom, the prominent visible structures are the 2.0 m primary mirror and support structure, the sunshields, the bus with radiative panels, the sunward-facing solar-panel skirt, and the high-gain antenna.  Figure reprinted from Fig. 3 of reference 23, with permission.} 
\end{figure} 

GEP (Fig. \ref{fig:GEPRendering}) is designed with a 2.0 m primary mirror that will be cooled to 6 K to enable sensitivity limited by photon shot noise from foreground astrophysical sources: zodiacal dust emission and Galactic dust emission. There is one scientific instrument with two modules: an imager, GEP-I, and a dispersive spectrometer, GEP-S. GEP-I has 23 photometric bands distributed on the focal plane: 18 resolution $R = \lambda/\Delta\lambda = 8$ bands from 10 to 95 $\mu$m designed to measure redshifts with PAHs and mid-infrared spectral energy distributions, and five resolution $R = 3.5$ bands from 95 to 400 $\mu$m to measure dust spectral energy distributions encompassing the peak to beyond $z = 2$. GEP-S utilizes four long-slit grating spectrometers with spectral resolution $R = 200$ from 24 to 193 $\mu$m. Both modules baseline arrays of kinetic inductance detectors (KIDs) cooled to 100 mK by a multistage adiabatic demagnetization refrigerator backed by a hybrid Joule-Thomson and Stirling cryocooler, which will also cool the telescope and coupling optics. GEP launch is targeted for January 1, 2029, with a planned mission duration of four years at Earth-Sun L2 (Table \ref{tab:GEPparameters}).

\begin{table}[ht]
\caption{Basic GEP parameters.} 
\label{tab:GEPparameters}
\footnotesize{
\begin{center}       
\begin{tabular}{|l|l|} 
\hline
\rule[-1ex]{0pt}{3.5ex}  Parameter & Quantity  \\
\hline\hline
\rule[-1ex]{0pt}{3.5ex}  Target Launch Date & January 1, 2029  \\
\hline
\rule[-1ex]{0pt}{3.5ex}  Orbit & Sun-Earth L2   \\
\hline
\rule[-1ex]{0pt}{3.5ex}  Observing Mode & Dedicated surveys  \\
\hline
\rule[-1ex]{0pt}{3.5ex}  Mission Duration & 4 years  \\
\hline
\rule[-1ex]{0pt}{3.5ex}  Telescope & 2.0 m, 4 K, unobscured, Au-coated SiC  \\
\hline
\rule[-1ex]{0pt}{3.5ex}  GEP-I Wavebands & 23 bands covering 10 - 400 $\mu$m  \\
\hline
\rule[-1ex]{0pt}{3.5ex}  GEP-I $R$ ($\lambda / \Delta\lambda$) & 8 (10 - 95 $\mu$m), 3.5 (95 - 400 $\mu$m)  \\
\hline
\rule[-1ex]{0pt}{3.5ex}  GEP-I Surveys and Target Depths & All sky, $\sim$ 1~mJy \\
\rule[-1ex]{0pt}{3.5ex} (obtainable with photon background-  & 300 sq. deg., $\sim50~\mu$Jy \\
\rule[-1ex]{0pt}{3.5ex} limited sensitivities) &  30 sq. deg., $\sim20~\mu$Jy \\
\rule[-1ex]{0pt}{3.5ex}  &  3 sq. deg., $\sim5~\mu$Jy \\
\hline
\rule[-1ex]{0pt}{3.5ex} GEP-S Bands &  24 - 42, 40 - 70, 66 - 116, 110 - 193 $\mu$m \\
\hline
\rule[-1ex]{0pt}{3.5ex}  GEP-S $R  (\lambda / \Delta\lambda)$ & 200  \\
\hline
\rule[-1ex]{0pt}{3.5ex}  GEP-S Surveys & Selected galaxies, 1.5 and 100 sq. deg.  \\
\hline
\end{tabular}
\end{center}
}
\end{table} 

\begin{figure}
\begin{center}
\begin{tabular}{c}
\includegraphics[height=5.5cm]{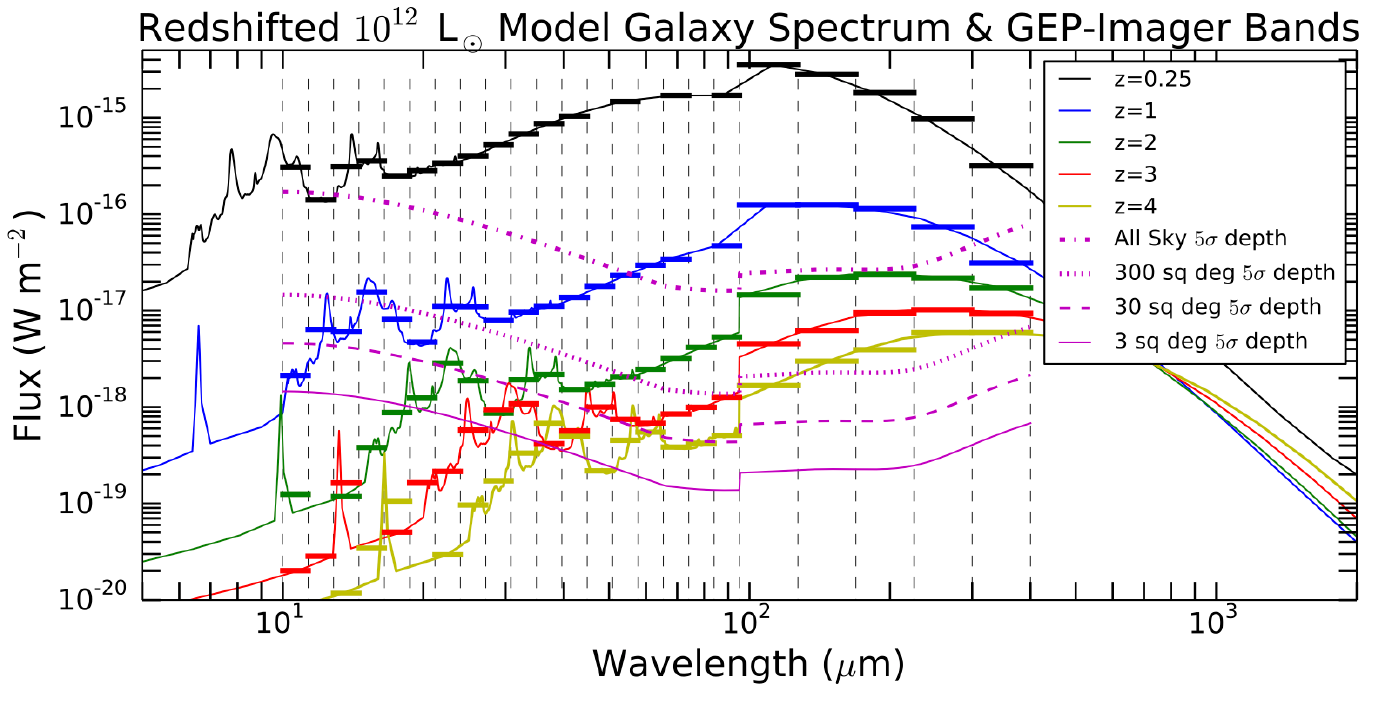}
\end{tabular}
\end{center}
\caption 
{ \label{fig:GEP_Model_Spectra}
GEP will detect $10^{12}$ L$_{\odot}$ galaxies at $z = 2$ and will measure redshifts with PAHs to at least $z = 4$ for bright galaxies and $z\approx7$ for gravitationally lensed galaxies. The spectra display PAH emission lines, silicate absorption at 10 $\mu$m (in the rest frame), a rising mid-infrared continuum from warm dust, and a peak just longward of 100 $\mu$m from cold dust. The spectra are binned into GEP-I’s wavebands, whose edges are demarcated by dashed vertical lines. The bandwidths change abruptly at 95 $\mu$m from $R = 8$ to $R = 3.5$ because the broad PAH emission lines are not expected to be bright enough for redshift determination for large numbers of galaxies. Atomic fine-structure emission lines and molecular lines are not shown. Magenta lines indicate 5$\sigma$ survey depths assuming photon background-limited sensitivities.} 
\end{figure}

GEP is optimized for large, multi-tiered surveys for galaxies detected by their mid- and far-infrared emission from dust, PAHs, and atomic fine-structure lines. The GEP reference mission includes two types of surveys: photometric hyperspectral surveys with GEP-I and spectroscopic surveys with GEP-S. The GEP-I survey areas will be 3, 30, and 300 square degrees, and an all-sky survey (Fig. \ref{fig:GEP_Model_Spectra}). 
The spectral surveys with $\lt1$\%-level redshift precision will cover a range of low- and high-ionization atomic fine-structure lines. Spectral surveys will consist of ‘blind’ surveys utilizing a long-slit configuration, intensity mapping, and follow-up, deep pointed observations of galaxies identified in the GEP-I surveys, and regions of the Milky Way and nearby galaxies.  GEP will require 4 years to do these surveys; however, it utilizes no expendable cryogens and a Guest Observer mission can be envisioned after the surveys are completed.

GEP shares common elements with two other cryogenic far-infrared space missions under study: SPICA, an ESA-JAXA collaboration in a similar cost class to GEP, and Origins Space Telescope (OST), a NASA concept study for a future flagship infrared mission. SPICA's design features a 2.5 m, 6–8 K telescope and a spectrometer using sensitive bolometer arrays,
although it does not have the 23-band, moderate spectral resolution hyperspectral imaging that is central to GEP’s architecture.
OST is designed with a 5.9 meter, 4.5 K telescope; its earliest possible realization is well into the 2030 decade. 

\subsection{This Paper}

GEP was described in the concept study report submitted to NASA  \cite{Glenn2019report} and its science case was briefly summarized in a short white paper submitted to the 2020 Astronomy Decadal Survey \cite{Glenn2019white}.  The initial design concept for the mission and the optical design were described in Glenn et al.\cite{glenn2018galaxy}, and the cryogenic design was described in Moore et al.\cite{moore2018thermal}.  This paper summarizes the design and science and provides new content: new engineering design, a summary of recent mid-infrared KID development, a discussion of microlens array fabrication for mid-infrared KIDs, and additional context for galaxy surveys.  

\section{Science Goals}


\label{sect:Goals}

GEP has two overarching science goals.  The first is map the history of galaxy growth through star formation and accretion by supermassive black holes and to characterize the relationship between these processes.  The second is to measure the growth of metals in galaxies and the changing of star-formation interstellar medium environments in galaxies over cosmic time.  These science goals are broken down into specific objectives, which are translated into requirements that drive the GEP design in the concept study science traceability matrix\cite{Glenn2019report}.  


Broadly, star formation began in the first billion years of the Universe, then rose to a peak or broad plateau around $ z = 2 - 3$, then declined sharply\cite{madau2014cosmic}. 
Despite substantial success over the past decades in understanding the average star formation and supermassive black hole accretion rates with redshift, large uncertainties and significant questions about galaxy evolution still exist, including:

star formation rates derived from infrared galaxy surveys of limited size have not probed low luminosities\cite{caputi2007} (essentially, below L$^*$) and are limited by sample variance, redshifts of large samples of far-infrared continuum-detected galaxies are uncertain, and extinction correction in rest-frame ultraviolet observations is substantial and sometimes uncertain\cite{madau2014cosmic}. GEP is designed to address these concerns with large, deep infrared surveys for star-forming galaxies to meet its first objective: to measure the coevolution of star formation and supermassive black hole growth in galaxies. 
GEP-I will yield star formation rates from far-infrared luminosities\cite{kennicutt1998global, battisti2015} and separate AGN and star formation contributions to total infrared luminosities via mid-infrared spectral shapes\cite{Dale2014}, even in cases of heavy extinction because the dust emission is almost always optically thin.
The 23-band GEP-I photometry of the mid-infrared PAH features will yield redshifts for the galaxies. PAHs have been detected spectroscopically with {\it Spitzer} at high redshifts, e.g., $z = 1.09$ and $2.96$\cite{teplitz2007measuring} and $z = 4.055$\cite{riechers2014polycyclic}, in galaxies with and without prominent AGN. 
From mid- and far-infrared fine-structure lines, GEP-S will measure star formation rates\cite{DeLooze2014,mordini2021calibration} and supermassive black hole accretion rates\cite{mordini2021calibration}.

GEP will access the early epochs of galaxy growth ($z > 3$) by utilizing the brightening from gravitational lensing. 
It is clear that much of the star formation and black hole growth in massive galaxies since re-ionization occurred in dusty regions and there is considerable uncertainty about how much star formation rate density may be missed in the ultraviolet census of that early phase of galaxy evolution\cite{madau2014cosmic}. Wide-area GEP-I surveys taking advantage of the brightening provided by gravitational lensing will address this problem with $10^4$ lensed high-z galaxies. 

Understanding the role of accreting supermassive black holes in galaxy evolution requires infrared observations 
because these processes occur on small scales at the centers of galaxies, in most cases obscured by the very gas and dust that is accreting. A principal objective of GEP is to identify obscured AGN in galaxies and relate their accretion luminosities to their star formation rates. GEP will identify and quantify luminosities of dust-obscured AGN via mid-infrared spectral signatures, including the continuum shape with GEP-I and high-ionization fine-structure atomic transitions with GEP-S, such as [Ne V] at rest-frame wavelengths of $14.3~\mu$m and $24.3~\mu$m and [O IV] at $25.9~\mu$m. 
GEP complements X-ray detection of AGN because X-ray observations can miss Compton-thick AGN or underestimate accretion rates.

The masses of supermassive black holes in the centers of modern-day galaxies are well correlated with galaxies' bulge masses \cite{magorrian1998demography,marconi2003relation}. This has led to the now commonly adopted hypothesis that a feedback loop exists in which AGN activity governs the rate of star-formation in massive galaxies, or at least in galactic bulges\cite{silk1998quasars}. Theoretical models\cite{bower2006breaking, croton2006many, sijacki2007unified, di2008direct} invoke AGN feedback as a primary mechanism to explain the observed distribution of galaxy masses today. Without AGN feedback, models are unable to explain the low ratio of galactic stellar mass to halo mass for high-mass galaxies\cite{benson2003shapes}. However, the efficiency of AGN feedback for regulating star formation has not been established empirically and remains controversial\cite{wagner2016galaxy, silk2013unleashing}.  GEP's goal is to determine how feedback from buried accreting black holes was related to the decline of star-formation.  
GEP will assess the role of feedback in galaxies
by searching for faint wings or asymmetries in the velocity profiles of bright fine-structure lines in stacked spectra of AGN and starburst samples.  Such line-wing signatures have been measured interferometrically in CO (e.g., Cicone et al. 2014 \cite{cicone14}) as well as in the optical (e.g., Vayne et al.\ 2021 \cite{Vayner21outflows}).  They can have velocities in excess of 1000 km/s, particularly for the AGN sources.  Given the modest spectral resolving power of GEP-S (R=200, or 1500 km/s), GEP's far-IR measurement will be accomplished with a combination of (a) high signal-to-noise ratio ($\sim$ 100 or greater) in the main line core for either the high-redshift spectral stacks or individual nearby galaxy spectra and (b) knowledge of the spectral response function of the instrument, with fidelity to 1 part in 500 or better.  The excellent raw sensitivity of GEP-S, combined with the fact that we are considering the brightest lines in the spectrum, will satisfy the first condition.  For the second, a combination of laboratory measurements pre-flight and measurements of narrow Milky Way line  line-emitting standards should provide ample opportunity for accurate spectral calibration.
In nearby galaxies, GEP will directly obtain a spatially-resolved view of feedback and its effects with a spectroscopic study of galactic outflows and fountains in local galaxies in various atomic fine-structure emission lines (e.g., [C II], [N II], and [O III]).  

GEP is designed not only to measure the star formation rates of galaxies over cosmic time, but also the evolving conditions of star formation regions.
Specifically, measurements of fine-structure lines, including [C II], [N II], and [O III], will be used to infer the mass and density of the interstellar medium gas and the hardness of the ambient radiation, which has implications for the stellar initial mass function, and the density of H II regions from which the interstellar pressure can be inferred.
The resulting relations between metallicity, star-formation rate, and other galaxy properties will inform models of galactic winds by placing constraints on the presence of gas-phase metals at their source.

Metallicity represents the integrated effects of star-formation, inflow, and outflow from galaxies. A key objective of GEP is to track the buildup of heavy elements in galaxies over the peak epoch of star-formation utilizing spectroscopic surveys of metals in atomic gas (and secondarily lower-resolution spectra of PAHs and dust).  
Metallicities of galaxies have not been measured beyond the local universe with the extinction-free probes afforded by far-infrared atomic fine-structure lines. GEP will measure the metallicity in galaxies down to L$^*$ $= 10^{12}$ L$_{\odot}$ galaxies at $z = 2$ using the nitrogen to oxygen ratio 
measured with [OIII] and [NIII] fine-structure lines.  This ratio measures metallicity through its connection to stellar processing since nitrogen is a secondary nucleosynthesis product and comes on later in the nucleosynthesis process.  The nitrogen / oxygen approach has been used with optical lines but the far-IR lines offer the dual benefit of dust immunity and temperature insensitivity. \cite{PereiraSantaella17O3N3}.  Alternatively, for the highest redshifts, metallicities will be measured using a diagnostic formed from the mid-IR tracers of neon and sulphur. This has been shown to track metallicity in photoionization models, and has been validated in a careful study of local galaxies which have Spitzer mid-IR spectroscopy in hand.  \cite{Fernandez21midIRmetallicity}

\section{Mission Architecture:  Thermal Design and Observing Modes}

GEP is designed as a Class B mission. 
A NASA Class B mission is a mission whose loss would be highly impactful to national science objectives and therefore that only low risks to mission success will be tolerated.  Less risk is tolerated for Class A missions and more for Classes C and D.  GEP is based on the Ball Aerospace BCP2000 reference bus, of {\it Kepler} heritage, customized to meet the science and mission requirements of GEP. The four-year survey program is divided into a GEP-I campaign and a GEP-S campaign, from which 350 TB of observational data will be downlinked. 
Mission details can be found in the concept study report\cite{Glenn2019report}.  

\begin{figure}
\begin{center}
\begin{tabular}{c}
\includegraphics[height=8.5cm]{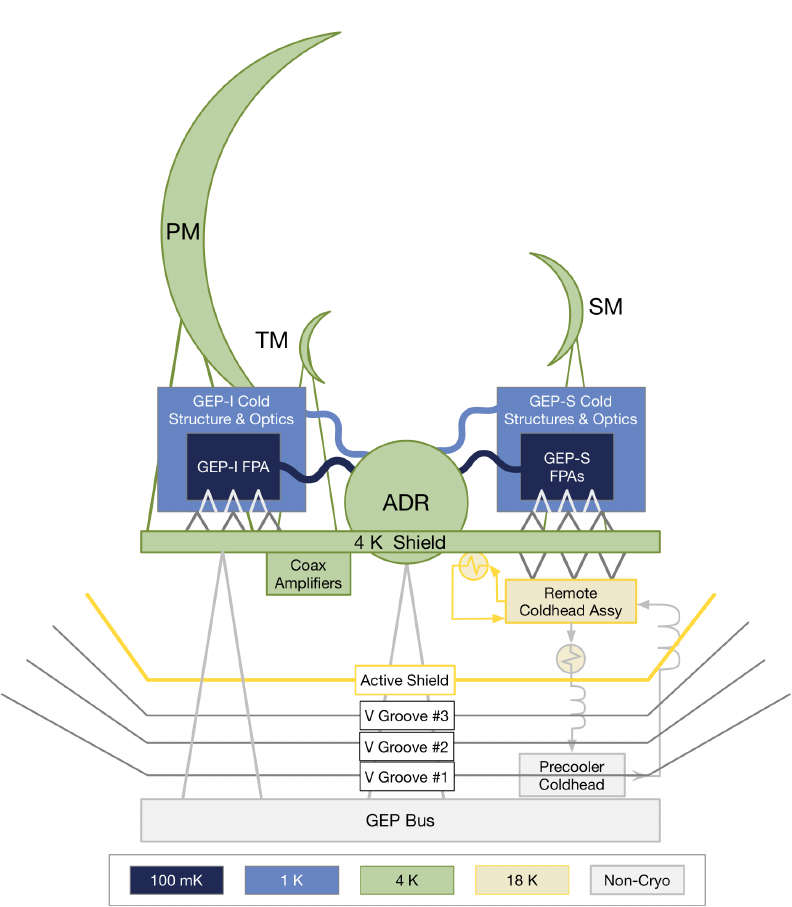}
\end{tabular}
\end{center}
\caption 
{ \label{fig:Thermal}
GEP thermal configuration.  GEPs thermal design uses active and passive cooling to maintain 100 mK required by the focal planes, 6 K for the optical assembly, including the telescope mirrors, and 18 K for the cryogenic amplifiers. This configuration supersedes the configuration in Moore et al. (2018)\cite{moore2018thermal}.  PM = primary mirror, SM = secondary mirror, TM = tertiary mirror, FPA = focal plane array.
} 
\end{figure}

GEP's thermal system employs multiple passive and active stages to meet the temperature intercept requirements of the instruments and optical assembly (Fig. \ref{fig:Thermal}).  A continuous multi-stage adiabatic demagnetization refrigerator (ADR) provides cooling for the detectors at 100 mK, with a 1 K thermal intercept to reduce thermal noise and parasitic loads. A hybrid Joule-Thomson/Stirling cryocooler intercepts heat at 4 K from the ADR and from the cryogenic amplifiers and parasitic loads. The hybrid cooler has an 18 K intercept to cool a second stage of amplifiers for the detector signals and an active shield. 

The sunshield assembly consists of three passively cooled reflective shields with a total area of 33 m$^2$, and a 9.77 m$^2$ active shield located above the sunshield under the focal plane boxes. The substrate of all four shields is an internally self-supporting 1 mm thick M55J laminate with an areal density of 29.5 kg m$^{-2}$. The reflective coating on the three warmer, passively cooled sunshields consists of a thin (0.005 inches) layer of aluminized Kapton adhered to the surface. The bottom surface of the actively cooled sunshield is also aluminized Kapton; however, the top surface is a 1 mm thick, high purity aluminum thermal spreader layer. Fiberglass composite struts provide intershield supports at the perimeter to accept launch loads.
The passively cooled shields intercept conducted loads from the bipods and harnesses and radiative loads from the Sun. With these non-deployable sunshields, the spacecraft can tilt $\pm 21.6^{\circ}$, maintaining all cryogenic components in the shadow cone and thus preserve thermal system operability. The cooler and ADR electronics dissipate heat at ambient temperature ($\sim 300$ K), along with all bus components, which is radiated by 10.5 m$^2$ of radiators mounted on the bus behind the solar panels.

The GEP-I and GEP-S instrument modules will observe one at a time and use the same readout electronics. They share a scan survey observing mode, where mapping is performed as the spacecraft slews at approximately 60\asec~s$^{-1}$. The scan survey mode will be used for all the GEP-I surveys, for GEP-S’s 1.5 and 100 square degree surveys, and for mapping of nearby galaxies. GEP-S will also have a pointed observation mode in which a chopping mirror, with a total throw of $\pm0.2^{\circ}$, modulates the signal for $1/f$ noise mitigation.

One five-hour DSN pass per day will be required to downlink 0.24 TB, during which the all-sky survey will be conducted by spinning the spacecraft while the high-gain antenna is pointed at Earth.  The intent is to do the entire all-sky survey in this mode while spinning at the same rate as the data taken in the nominal scan mode.  If it is determined that the scan rate must be different for the all-sky mode than for nominal scanning, a separate calibration will be required.


\section{Science Instrument}

\label{sect:Instrument}

\begin{figure}
\begin{center}
\begin{tabular}{c}
\includegraphics[height=5.5cm]{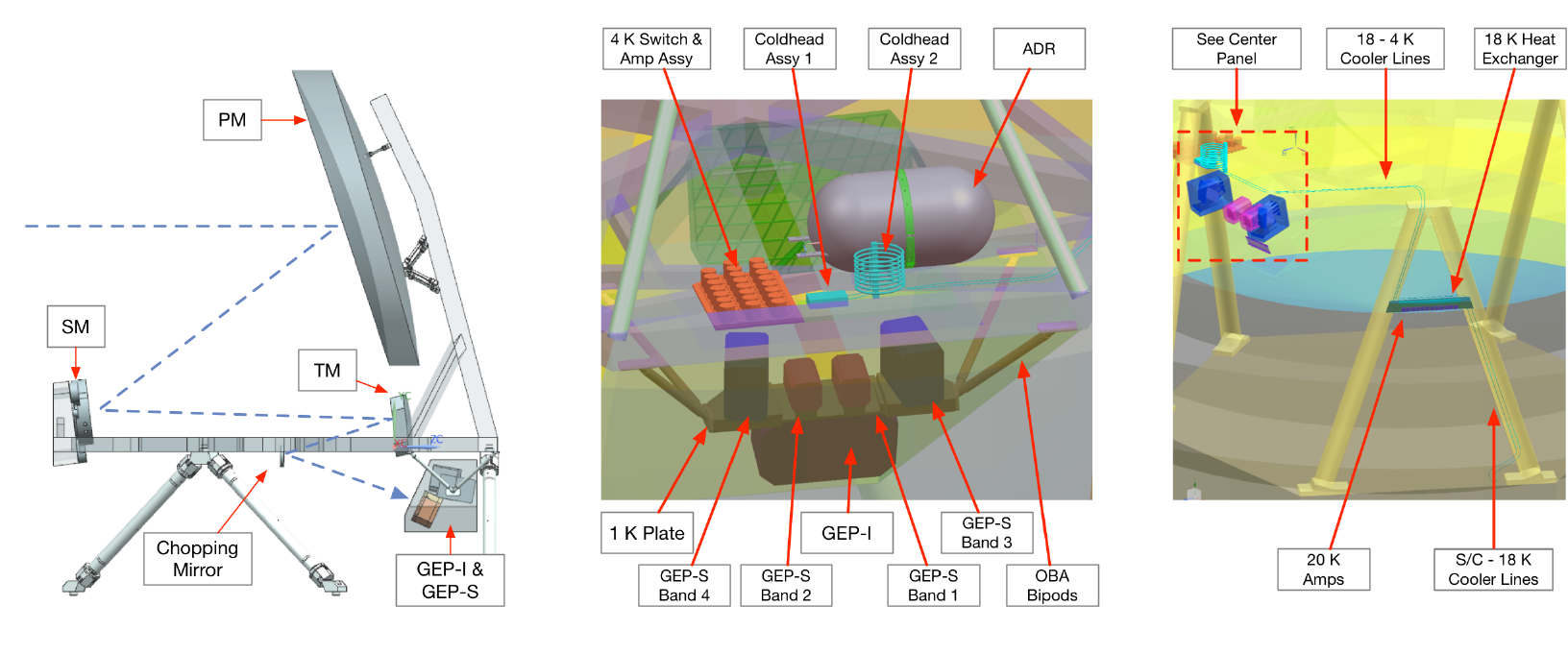}
\\
\hspace{1.0cm} (a) \hspace{4.0cm} (b) \hspace{3.5cm} (c)
\end{tabular}
\end{center}
\caption 
{ \label{fig:Payload}
(a) Side view of the GEP optical configuration showing its three-mirror astigmat design. A field stop is located between the secondary and tertiary mirrors. The fourth optic is a chopping mirror located at an image of the primary mirror, forming a pupil stop. The GEP-I and GEP-S modules are in the lower right. (b) Detail of GEP-I placement and the four GEP-S focal planes along with the ADR and 4 K cryogenic assembly.  Reprinted from Fig. 11 of reference 23, with modifications and permission. (c) Detail of the 18 K assembly located near the focal plane assembly.
} 
\end{figure}

The GEP payload includes the optical telescope assembly, the GEP instrument comprised of a hyper-spectral imager module (GEP-I) and four spectrometer modules (GEP-S), and a payload thermal subsystem (Fig. \ref{fig:Payload}). 

\subsection{Optical Design}

\label{sect:Optics}

The fundamental optical design requirement is to collect light with a 2.0 m diameter primary mirror and to form an f/9 focus. An unobscured three-mirror astigmat (TMA) is an oft-used configuration that is well suited to the first-order optical and mechanical requirements of this system. The powered mirrors are all conic shapes with parent surfaces that have mutual tilts and decentrations to reduce wavefront error.  The baseline for the GEP primary, secondary, and tertiary, and chopping mirrors is unactuated silicon carbide (SiC), as was flown on the {\it Herschel Space Observatory} and {\it Gaia}. 
GEP's optical design and simulated performance are described in detail in Glenn et al.\cite{glenn2018galaxy}. 

The five instrument interface planes are nearly coplanar at the common focal surface of the TMA, with minor differences in final focus to minimize wavefront error in each channel. Based on the sizes of the focal-plane array and the spectrometer slits and enclosures, the optimized field of view is 
$0.81^{\circ} \times 0.88^{\circ}$, in excess of what is required for the instrument modules. The centers of the GEP-S slits are in the plane of symmetry of the TMA so that they can be untilted with respect to the central ray to each spectrometer. Stray light suppression is achieved with a pupil stop and a field stop.

To meet the extragalactic confusion requirement (section \ref{sect:Confusion}), the primary mirror is required to be diffraction limited at $\lambda = 24~\mu$m. However, the root-mean-square wavefront error across the field of view in the optical design is diffraction limited at the minimum wavelength of $10~\mu$m. The 100\% encircled energy diameter is $\le 1.45$\asec, much less than the 3.43\asec~detector pixel size (see Section
\ref{sect:GEP-I}).  

\subsection{GEP-I Hyperspectral Imager}

\label{sect:GEP-I}

\begin{figure}
\begin{center}
\begin{tabular}{c}
\includegraphics[height=4.5cm]{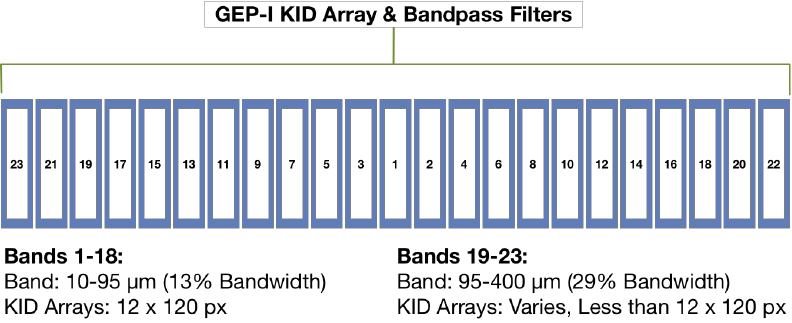}
\end{tabular}
\end{center}
\caption 
{ \label{fig:FocalPlane}
GEP-I focal plane layout configuration. While the entire focal plane shown is diffraction limited at $\lambda = 10~\mu$m, the shortest wavebands are located at the center where the optical quality is the highest.  Bands 1 ($10.0-11.3$ $\mu$m) through 15 ($57.7-65.4$ $\mu$m) have 1,440 KIDs (arrays of $12\times120$), with 3.43\asec~ pixels (300 $\mu$m square). The longer-wavelength bands have fewer, larger KIDs such that the focal plane area occupied by each band is approximately the same. For example, Band 23 ($300–400$ $\mu$m) has 46 KIDs 1,560 $\mu$m square. 
} 
\end{figure}

GEP-I is designed to obtain repeated measurements in each of the 23 wavebands to build survey depth by continuously scanning areas of sky covered by each survey. Each waveband occupies the same amount of focal plane area, 0.002 square degrees, half of which is occupied by detectors and half of which is allotted for bandpass filter mounting in the current design configuration(Figs. \ref{fig:FocalPlane}). 
Bands 1–18 have spectral resolution $R = \lambda / \Delta\lambda = 8$, whereas bands 19–23 are $R = 3.5$.  It is likely that the mid-infrared (e.g., $10 - 26~\mu$m) bandpass filters could be replaced with linear-variable filters, resulting a spectral resolution of at least $R = 20$, which will be the subject of a future investigation.

GEP-I will have 25,735 KIDs. GEP I’s optical design performance is diffraction limited at $10~\mu$m; however, the primary mirror is specified to be diffraction limited at $24~\mu$m, corresponding to $\sim3$\asec~(FWHM) beam size. The primary reason for this is that the smallest KID pixel size we expect to be able to fabricate without exceeding the readout bandwidth  is $300~\mu$m $\times$ $300~\mu$m (the number of detectors is inversely proportional to their physical area), which corresponds to 3.43\asec. 
With 3\asec~beams and 3.43\asec~ pixels, the full angular resolution of the telescope will not be taken advantage of for GEP-I bands 1 - 9 (up to $29~\mu\mathrm{m}$), although all flux density will be recovered.  For bands 9 - 23, the pixel size will be less than the beam FWHM.  Should greater bandwidth become feasible through improvements in data acquisition and computing speed (a very likely development), the pixel sizes can be reduced to less than the beam FWHM.

\subsection{GEP-S Long-Slit Spectrometer}

\label{sect:GEP-S}

GEP’s spectrometer was designed to meet the science requirements calling for observing mid- and far-infrared atomic fine-structure lines from galaxies over a range of redshifts. Specifically, the $24.3~\mu$m [Ne V] line starting at $z = 0$ for AGN identification and the $63.2~\mu$m [O I] line at 
$z = 2$ for probing star-forming galaxies at or beyond the cosmic peak. The entire bandwidth should be available to identify spectral lines for galaxies of unknown redshift. Sufficient spectral resolution is required to achieve good sensitivity through dispersion of the astrophysical background photons. Spectral resolution $R = 200$ meets these requirements.

\begin{figure}
\begin{center}
\begin{tabular}{c}
\includegraphics[height=10.5cm]{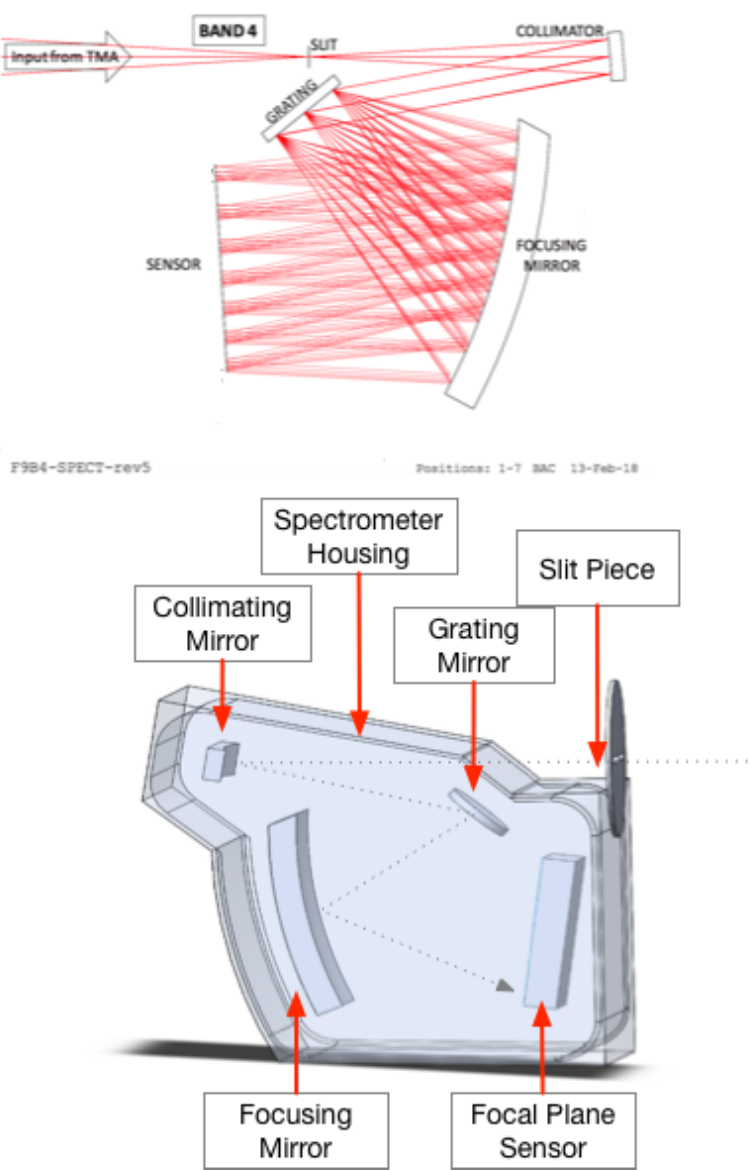}
\end{tabular}
\end{center}
\caption 
{ \label{fig:Spectrometer}
(Top) Ray-trace of GEP-S's Band 4 spectrometer module.  Reprinted from Fig. 6 of reference 23, with permission. (Bottom) GEP-S Band 4 mechanical layout viewed mirror imaged left-to-right with respect to the top figure.
} 
\end{figure}

GEP-S's design is implemented with four diffraction gratings, each with fractional bandwidth of 1.6, identified as bands 1–4 (Table \ref{tab:GEPparameters}), with a total of 24,640 KIDs. The ray trace diagram and CAD model of band 4 is shown in Fig. \ref{fig:Spectrometer}. Each band, or diffraction grating module, is comprised of an enclosure, a slit, a collimator, a grating operated in first order, a focusing mirror, and an array of KIDs. The long slit lengths enable spectral mapping. As with GEP-I, the shortest-wavelength bands 1 and 2 are placed nearest to the center of the field of view where the optical performance is the best.  A chopping mirror (Fig. \ref{fig:Payload}, panel (a)) is included for detector modulation for staring GEP-S observations.

\subsection{Kinetic Inductance Detectors and Readout Electronics}

\label{sect:techdev}

\begin{figure}
\begin{center}
\begin{tabular}{c}
\includegraphics[height=7.0cm]{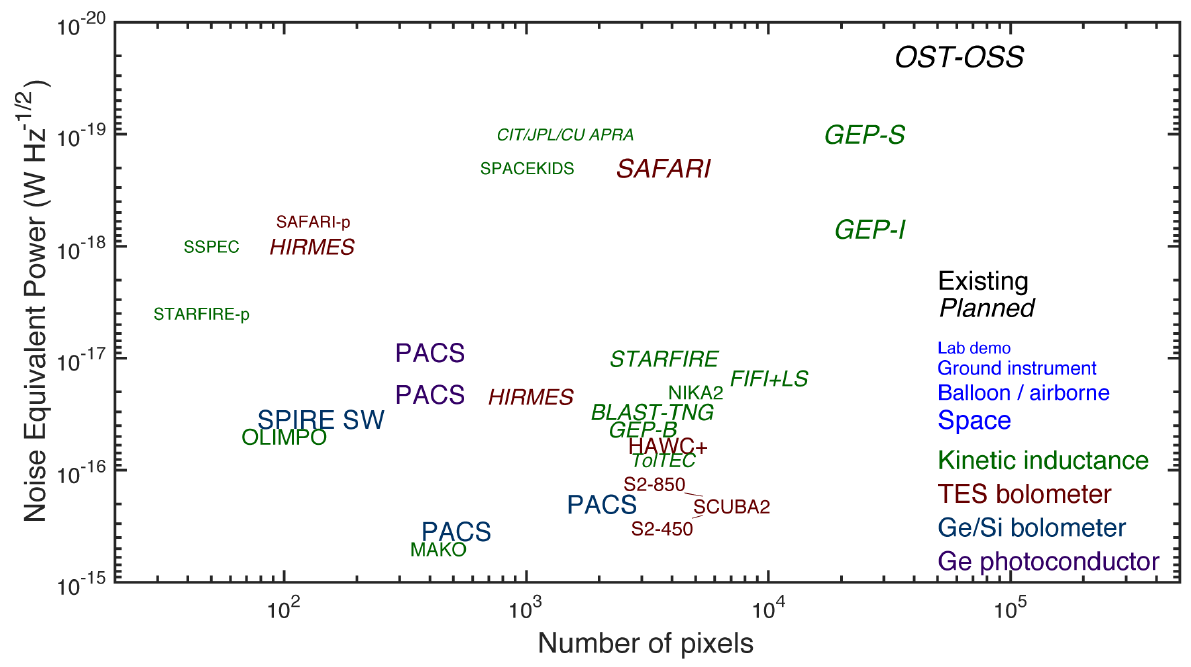}
\end{tabular}
\end{center}
\caption 
{ \label{fig:FIRTech}
Far-IR detector technology has made considerable progress toward larger, more sensitive arrays over the past decade. Detector sensitivity is shown versus numbers of detectors for existing (normal font) and planned (italic font) instrumentation. The text colors indicate the type of detector technology and the font sizes are used to differentiate the testing environment or intended application of the technology. GEP-I sensitivity requirements have been met by SPACEKIDs\cite{baselmans2017kilo}, while GEP-S will require further sensitivity improvement for background-limited observations. Both GEP-I and GEP-S require KIDs at mid-infrared wavelengths, for which development efforts are underway\cite{hailey2018development}. Others plotted are BLAST-TNG\cite{lourie2018preflight}, FIFI+LS (proposed), 
HIRMES\cite{nikola2018hirmes}, MAKO\cite{swenson2012mako, mckenney2012design}, NIKA2\cite{adam2018nika2}, OLIMPO\cite{paiella2019kinetic}, OST-OSS\cite{bradford2018origins}, PACS\cite{poglitsch2010photodetector}, SAFARI-p\cite{hijmering2016readout},  SAFARI\cite{de2018safari}, SCUBA2\cite{holland2013scuba},   SPIRE\cite{griffin2010herschel},  SuperSpec\cite{wheeler2018superspec},  STARFIRE-p\cite{barlis2018development, hailey2018development}, STARFIRE\cite{aguirre2018starfire}, and TolTEC\cite{wilson2018toltec}. In addition to its own scientific program, GEP fills an important role between the detector technology achievable in the next few years and possible future flagships such as OST.
} 
\end{figure}

KIDs have been baselined for the GEP-I and GEP-S focal plane arrays.  KIDs are hiqh quality factor superconducting resonators whose resonances vary with absorbed light through modulation of kinetic inductance by pair breaking\cite{day2003broadband}.  They are in wide use in recent and imminent ground-based and suborbital optical, far-infrared, submillimeter, and millimeter-wave instruments (Fig. \ref{fig:FIRTech}).  

Two other detector array technologies were considered for GEP:  transition-edge sensors (TESs) and Si:As IBCs (Impurity Band Conductors).  KIDs are not as technologically mature as TESs but their simple fabrication and focal plane electronics make them advantageous, partially because TESs require complex hybridization with SQUID readouts.  Like KIDs, TESs have not been demonstrated in the 10 -- 15 \micron~wavelength regime.  Si:As IBCs are technologically mature but they are unusable beyond 28 \micron, so KIDs (or TESs) would still be needed for 28 - 400 \micron. However, pixel pitches and operating temperature requirements for IBCs (7 K) and KIDs (100 mK) are very different, and the divergent optical and cryogenic requirements would effectively necessitate two instruments each for GEP-I and GEP-S, which would drive up complexity, cost, and risk.  Thus, KIDs have been adopted as the baseline detector technology for all of the wavebands of GEP-I and GEP-S.


GEP-I and GEP-S will utilize back-illuminated, lumped-element, microlens-coupled aluminum KIDs. Each KID is a superconducting thin-film microresonator, comprised of an absorbing meandered inductor and an interdigitated capacitor deposited on a silicon substrate. Light is concentrated onto the inductors by a microlens array. The geometry of the inductor controls the absorption characteristics as a function of wavelength and polarization. Absorption of photons changes the kinetic inductance, producing a small shift in the resonant frequency and resonance depth, which are detected in the transmitted probe signal. To enable frequency-multiplexed readout with a minimal bandwidth requirement, the interdigitated capacitors are unique for each pixel, yielding different resonant frequencies separated by several resonance line widths. Submillimeter-wavelength KIDs with sensitivity sufficient for GEP-I have been demonstrated  with a different architecture\cite{baselmans2017kilo}.  NEP improvement, not just different operational wavelengths, is needed for GEP-S.

GEP KIDs are based on the MAKO design\cite{mckenney2012design} and adapted to work at shorter wavelengths.  The MAKO design will work down to a wavelength of approximately 100 \micron, but below that modification will be required: the wire grid absorber linewidth used in the 350 \micron~MAKO detectors is 0.4 \micron.  Scaling this design with wavelength would require prohibitively narrow lines for $\lambda \lt 100$ \micron, even if using an ultraviolet stepper for film deposition. Our approach is to interrupt the absorber line with a meander that increases the resistance per unit length (and therefore the absorption efficiency), while simultaneously introducing a distributed capacitance to compensate for the accompanying inductance to tune the impedance \cite{Perido2020}. An impedance match to silicon may then be maintained with much wider and easier-to-fabricate lines. 
 
\begin{figure}
\begin{center}
\begin{tabular}{c}
\includegraphics[height=5.0cm]{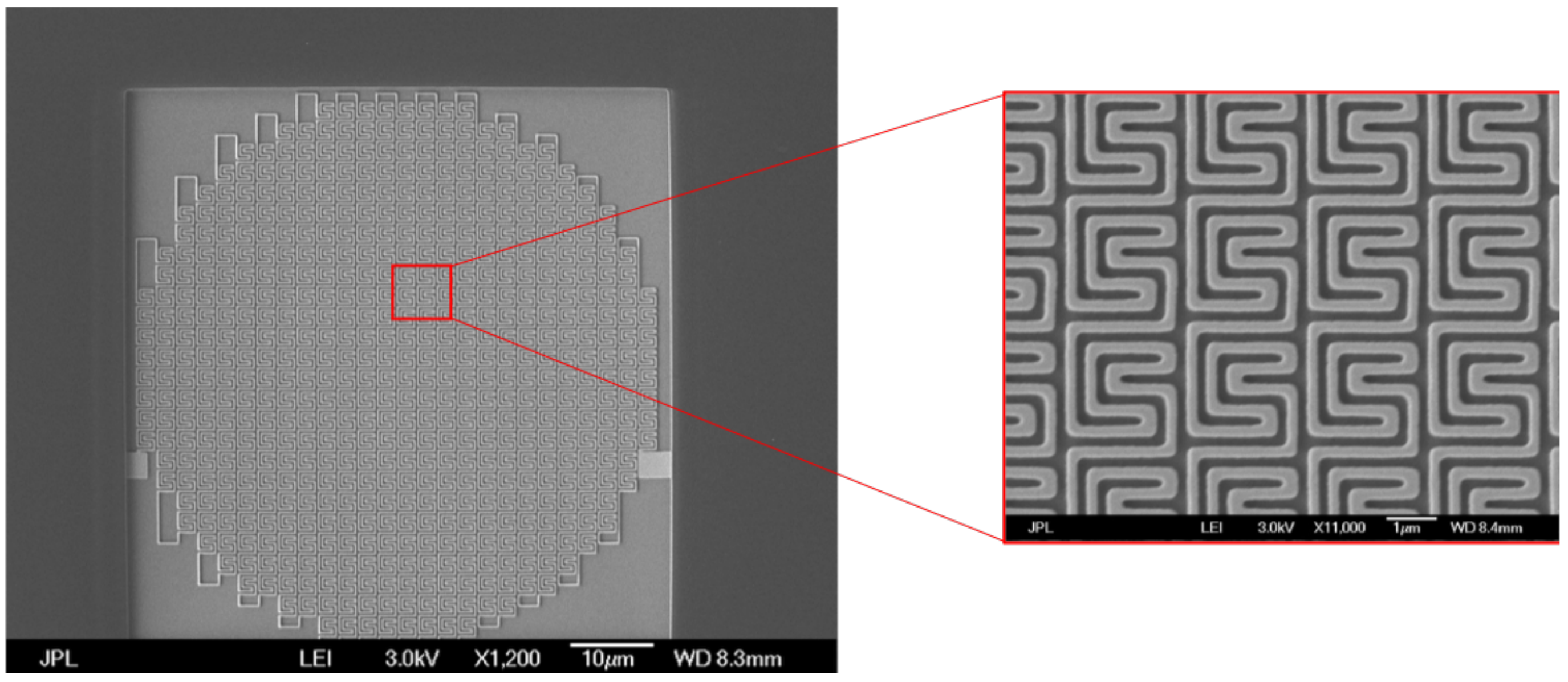}
\\
(a) \hspace{4.5cm} (b)
\end{tabular}
\end{center}
\caption 
{ \label{fig:Fab_3D_Combo_part}
A photograph the absorptive portion of a $\lambda = 10~\mu$m KID, courtesy H. G. LeDuc, reprinted by permission of the Creative Common license (creative commons.org/licenses/by/4.0/legalcode) from Fig. 1 of reference 63. (a) The inductor/absorber portion of the KID, which has a diameter of 60 $\mu$m.  The unit cell size is $2.4~\mu\mbox{m} \times 2.4~\mu$m. The inductor is coupled to a larger interdigitated capacitor to form a microresonator.  (b) Approximately 4 (horizontally) by 3 (vertically) unit cells. This absorber absorbs the vertical sense of polarization. }
 
\end{figure}

This absorber design is shown in Fig. \ref{fig:Fab_3D_Combo_part}. For 10 \micron\ radiation, the linewidth is 200 nm and the unit cell is 2.4 \micron\ square, repeated continuously in the circular envelope of the absorber with electrical continuity meandered vertically. Simulations indicate that this geometry achieves a peak single-polarization absorptivity of just over 70\% near 10 \micron\ for $R_\square = 0.2 - 0.8~\Omega$, bracketing the range expected for a 40 nm Al film.   Perido et al.\cite{Perido2020} present dark measurements of the KID array shown in Figure \ref{fig:Fab_3D_Combo_part}. 
The frequency noise was dominated by two-level system noise, with typical values at 10 Hz of $S_{xx} = (3-0.6) \times 10^{-17}$ \sxxunits\ at $T = 100-200$ mK, consistent with the expected $T^{-1.7}$ scaling.  Perido et al.\cite{Perido2020} also presented a resonant dual-polarization design for which simulations yield absorption of 70\% in each polarization. 

The pixel-to-pixel spacing is 300 \micron, dominated in area by the resonators' interdigitated capacitors.  Efficiently coupling the inductive absorbers to free space radiation will be achieved with microlens arrays -- large feedhorn arrays are currently impractical for short mid-infrared wavelengths because manufacturing tolerances are too small.
For minimal absorption by the substrates, the arrays should be fabricated with silicon for wavelengths greater than 20 \micron\ and germanium for 10 - 20 \micron\ because of the silicon absorption features in the $14 - 16$ \micron\ range.  Microlens arrays could either be fabricated separately and bonded to back-illuminated KID arrays \cite{Defrance2018} or etched into the wafer frontside before or after KID fabrication.  There are examples of silicon microlens fabrication in the literature that could achieve the micron-level tolerances required for operation down to $10$ \micron \cite{Savander1994,Larsen2005,Chenetal2005,Belmonteetal2014,Dengetal2015,Mengetal2015,Zuoetal2017}, although those involving micromachining and laser etching would likely be too slow for arrays of $10^4$ detectors. 
A promising approach is the deposition of Fresnel 'zone plate' microlens arrays on the wafer frontsides, which have already been demonstrated at $\lambda = 10.6$ \micron\ for antenna coupling\cite{Gonzalez2004}.  This approach has the virtue of fabrication simplicity but it rejects 50\% of radiation. If noise-equivalent powers (NEPs) must be improved by a factor of $\sqrt{2}$, three-dimensional hemispherical lenses or Fresnel lenses could be etched on the wafer frontsides instead of depositing Fresnel zones.

\begin{figure}
\begin{center}
\begin{tabular}{c}
\includegraphics[height=7.5cm]{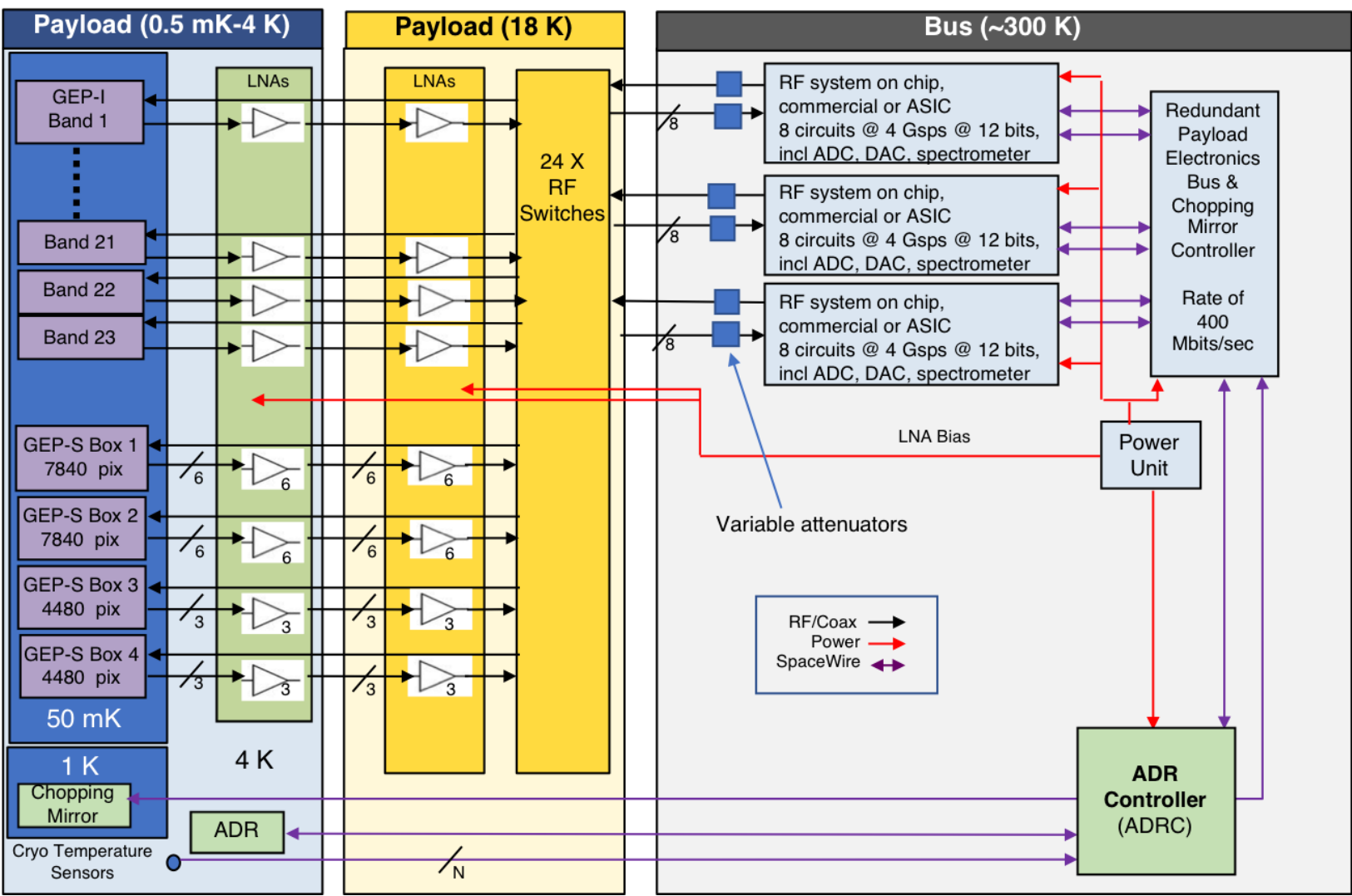}
\end{tabular}
\end{center}
\caption 
{ \label{fig:Readout}
Readout system for the GEP KID arrays. Electronics will be room temperature ($\sim300$ K) except the amplifiers at 4 K and 18 K and the KIDs themselves at 0.1 K. The power consumption is estimated at 25 W per readout circuit. GEP-I uses twenty-three readout channels, GEP-S Bands 1 and 2 use six readout channels each, and GEP-S Bands 3 and 4 use three readout channels each.
} 
\end{figure}

For all GEP focal planes, KIDs are organized into groups of approximately 1,500 detectors that are spread
across a 0.6 - 1.6 GHz band and read out using electronics illustrated in Fig. \ref{fig:Readout}. The choice of a 1.1 GHz center frequency resulted from a trade study in which smaller pixels were favored by the optical design but larger pixels reduced the readout frequency and bandwidth, and thus power dissipation. The readout electronics generate analog waveforms using RF-DACs that are transmitted to the cold focal plane, exciting all 1,500 resonators. The 1 GHz bandwidth return signal from all 1,500 KIDs is digitized with RF-ADCs and digitally channelized with sufficient resolution to separate the individual KID frequencies. 
The KID readout scheme, initially demonstrated in 2006\cite{mazin2006digital}, has now been implemented in various forms for ground-based and balloon-borne instruments\cite{yates2009fast, duan2010open, duan2015, mchugh2012readout, swenson2012mako, strader2016digitial, bourrion2016nikel_amc, van2016multiplexed, gordon2016open, henderson2018highly}. 

The GEP-I and GEP-S modules will share the same
set of readout electronics (Fig. \ref{fig:Readout}): they have separate, non-simultaneous observing modes. Microwave switches on the readout lines enable selection of GEP-S or GEP-I for readout. There are 24 parallel readout channels available from three RF readout boards with eight channels each. 
The eight-channel RF readout cards share a payload electronics chassis with a chopping mirror driver card and two clock, processor, memory, and power cards for dual-string redundancy. The total estimated power consumption for the readout electronics is 484 W.

\subsection{KID Development and Readout Outlook}

For a launch date as early as January 1, 2029, technologies required for GEP must be at or above TRL 6 in 2025.  Mid-infrared KIDs are currently at TRL 3, with the proof-of-concepts having been demonstrated.  To reach TRL 6, they must be validated at a component level in a relevant environment and in a subsystem operational environment.  In this case, the latter indicates integration with optics and readout electronics.  Here, we review the current state of the art and itemize the technological improvements that must be made at a component level.  This will have to followed by demonstrations of sensitivity, yield, and sufficiently low cosmic ray susceptibility for arrays while read out with flight-like electronics.  Of the existing demonstrations shown in Fig. \ref{fig:FIRTech}, the SPACEKIDs results\cite{baselmans2017kilo} are the closest to meeting the GEP requirements, which are listed in Table \ref{tab:DetReqGEPvsSpace}. By comparing the capability gap between GEP KID focal plane requirements and SPACEKIDs performance, specific advances required to achieve TRL 6 can be identified.

\begin{table}[ht]
\caption{Detector Requirements for GEP versus achieved for SPACEKIDs \cite{baselmans2017kilo}.} 
\label{tab:DetReqGEPvsSpace}
\scriptsize{
\begin{center}  
\begin{tabular}{|c|c|c|c|c|c|c|c|c|c|c|c|} 
\hline
\rule[-1ex]{0pt}{3.5ex}   & \shortstack{Tile Size \\ (pixels)} & \shortstack{$\lambda$ \\ ($\mu$m)} & \shortstack{NEP \\ (W Hz\textsuperscript{-1/2})} & \shortstack{MUX \\ (pix/GHz)} & \shortstack{Pitch \\ ($\mu$m)} & \shortstack{$\tau_{det}$ \\ (ms)} & \shortstack{Min. \\ Yield} & \shortstack{Dynamic \\ Range} & \shortstack{Crosstalk \\ (dB)} & \shortstack{1/f knee \\ (Hz)} & \shortstack{Cosmic \\ Ray \\ Deadtime} \\
\hline\hline
\rule[-1ex]{0pt}{3.5ex}  SPACEKIDs & 961 & 350 & 3 x 10\textsuperscript{-19} & 240 & 1600 & 1.5 & 83\% & 10\textsuperscript{5} & -30 & 0.2 & 5\%  \\
\hline
\rule[-1ex]{0pt}{3.5ex}  GEP-I & 1,440 & 10-400 & 7 x 10\textsuperscript{-19} & 1,500 & 300 & \textless4 & 80\% & 5,000 & -17 & \textless1 & \textless2\% \\
\hline
\rule[-1ex]{0pt}{3.5ex}  GEP-S & 980 & 24-193 & 1 x 10\textsuperscript{-19} & 1,500 & 300x600 & \textless4 & 80\% & 1,000 & -17 & \textless1 & \textless2\%  \\
\hline

\end{tabular}
\end{center}
}
\small
Notes: Minimum tile sizes for GEP-I/GEP-S shown; actual arrays could be multiples thereof. Tiles with 12 × 120 = 1,440 pixel format are envisioned for GEP-I bands 1–18. GEP-S bands 1 \& 2 assume arrays with 112 × 70 format, which could consist of tiles with 28 × 35 = 980 pixels. The dynamic range is specified for a 1 Jy calibration source, e.g., an asteroid \cite{muller2014herschel, baselmans2017kilo}. Techniques to mitigate electrical and optical crosstalk, and cosmic ray susceptibility, have been demonstrated\cite{noroozian2012crosstalk, baselmans2017kilo, yates2017surface}.
\end{table} 

\begin{enumerate}

\item
Sensitivity: To be astrophysically photon background limited, which results in the best possible sensitivity, the telescope and optics must be below 6 K and the detector noise-equivalent powers (NEPs) must be below the quadrature sum of all the other NEP terms (Fig. \ref{fig:AstrophysicalBackgrounds_Corrected_6K}):  $7 \times 10^{-19}$ \nepunits\ for GEP-I and $1 \times 10^{-19}$ \nepunits\ for GEP-S. GEP-S requires at least a factor of 3 improvement in NEP over SPACEKIDs. This can be achieved through reducing the detector active volume (thereby increasing the quasiparticle density and responsivity) below the $\sim100~\mu^3$ used by SPACEKIDs.  

\begin{figure}
\begin{center}
\begin{tabular}{c}
\includegraphics[height=5.5cm]{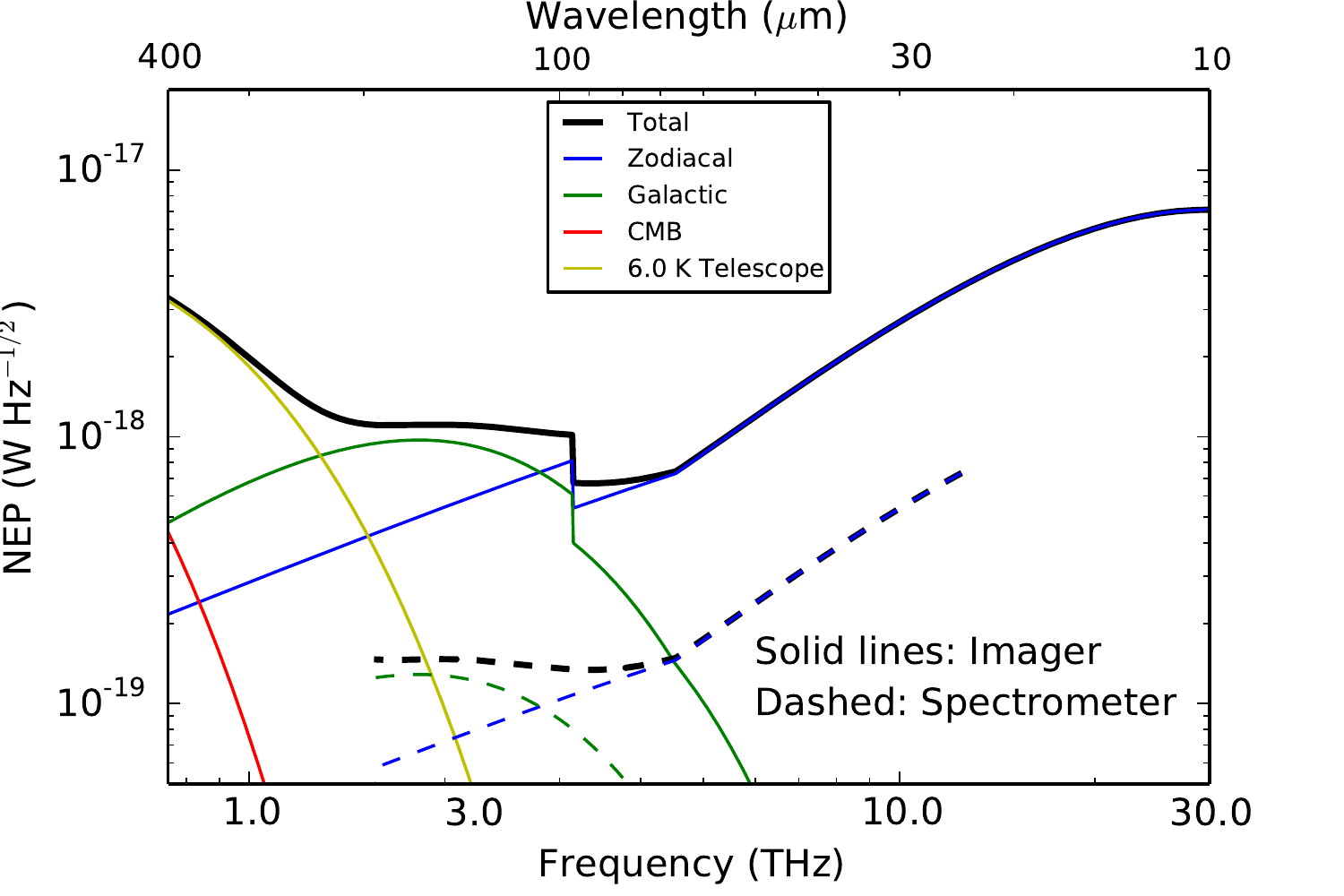}
\end{tabular}
\end{center}
\caption 
{\label{fig:AstrophysicalBackgrounds_Corrected_6K}
Detector NEPs for various photon backgrounds for the zodiacal light, the Galactic thermal dust emission, and the GEP telescope and instrument parameters.  This  shows why a $T_{tele} \le 6$ K telescope is necessary:  it is not possible to achieve astrophysical photon background-limited performance for wavelengths longer than 200  $\mu$m with $T_{tele} \gt 6$ K. A cooler telescope would have better GEP-I sensitivity, but it would be of limited value because the astrophysical source confusion is expected to be severe for $\lambda \ge 200~\mu$m.  GEP-S would benefit from cooling the telescope to 4 K even at the longest wavelengths, so this possibility may be adopted later.} 
\end{figure}

\item
Multiplexing: The minimum detector-multiplexing factor for GEP is 1,500 pixels per GHz of readout bandwidth. This $6\times$ increase in multiplex factor relative to SPACEKIDs is based on a $6\times$ reduction in readout frequency, from 6 GHz to 1 GHz (GEP's microwave readout band center). Multiplex factors of 6,500 per GHz have been demonstrated at 200 MHz readout frequency\cite{swenson2012mako}. Further improvements have been demonstrated using post-fabrication resonator trimming methods\cite{liu2017superconducting, shu2018increased}.

\item
Pitch: The baseline GEP-I pixel pitch is 300 \micron\ from wavelengths of 10 \micron\ to 60 \micron\ to provide 3.43\asec\ pixel sizes (the pitch is larger for $\lambda \gt 60~\mu$m). For GEP-S, the pixel pitch is $300 \times 600$ \micron\ to meet the spectral resolution requirement. 
The GEP requirements can be met with the design in Fig. \ref{fig:Fab_3D_Combo_part} and microlens arrays. Smaller pixels have smaller inductors and area-dominating interdigitated capacitors, and therefore higher resonance frequencies, which presents an engineering design trade space of pitch versus readout frequency.

\end{enumerate}

\begin{figure}
\begin{center}
\begin{tabular}{c}
\includegraphics[height=5.5cm]{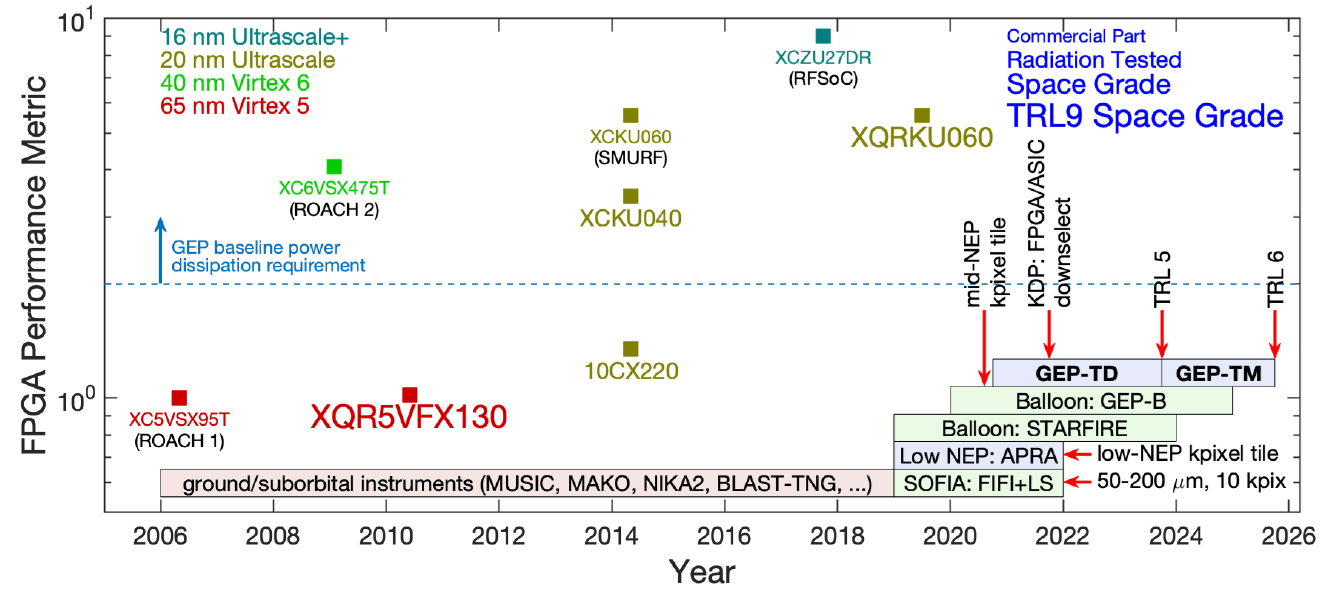}
\end{tabular}
\end{center}
\caption 
{ \label{fig:FPGA}
As part of the GEP concept design study, a plan was developed to mature KIDs and payload electronics technologies in advance of an anticipated Probe-class proposal through laboratory (shaded in red), SOFIA and balloon-borne instruments (green), and system-level demonstrations (blue) that started in 2019. 
In the top of the figure, feasibility of GEP readout electronics is illustrated using the evolution of Xilinx FPGA capability over time. Specific devices are plotted as points labeled by the part number, with data sheets listed in the references. The horizontal axis gives the introduction date, and the vertical axis is a composite performance metric that combines the number of logic gates and DSP cells, memory, and IO capability. The dashed blue line indicates the approximate minimum capability required for meeting the GEP baseline power dissipation of 25 W per 1 GHz readout channel. Color is used to indicate the CMOS technology node while font size indicates suitability for use in the space radiation environment. The XQRKU060, which Xilinx plans to release as a space-grade part in 2020, is a strong candidate. GEP's technology development plan calls for a TRL 6 validation of all hardware required for GEP’s payload by 2025. FPGA data sheets are listed in references \cite{xilinx2013,xilinx2018,xilinx2019,xilinx2020,xilinxvirtex52015,xilinxvirtex62015}
} 
\end{figure}

To date, all KID instrument readout systems use field-programmable gate arrays (FPGAs)\cite{Trimberger2015} for the digital channelization. This channelization is conceptually equivalent to an FFT, and sometimes is implemented as such\cite{yates2009fast, van2016multiplexed}. 
First-generation KID readouts\cite{duan2010open, duan2015, mchugh2012readout, swenson2012mako} typically used Virtex-5\cite{ROACH1-2018} hardware to process a 500 MHz bandwidth using about 50 W, across which up to 4,000 KIDs could be multiplexed\cite{swenson2012mako}. Similarly, the CORE mission study\cite{de2018exploring} concluded that a 1 GHz bandwidth could be read with 50 W using existing space-grade, TRL-9 Virtex-5 FPGAs\cite{elftmann2018}.  
Newer-generation FPGAs could have $10\times$ lower power consumption\cite{xilinxvirtex62015}, (Fig. \ref{fig:FPGA}), thus the laboratory figure of 50 W per 1 GHz channel is conservative, and there is strong interest in the use of advanced FPGAs in space\cite{wirthlin2013fpgas, swift2017invited, SEE2018, SEFUW2018, le2018efpga, wang2018live} 
Radiation test results on late-generation FPGAs have been positive\cite{lee2015single, lee2017commercial, elftmann2018, AllenandVartanian2018}. As a result, Xilinx's  20 nm KU060 FPGA (used in the SMURF readout electronics developed at SLAC)\cite{henderson2018highly} will be available as a space-rated product by late 2020\cite{elftmann2018}. This bodes well for even more advanced options, such as the new 16 nm FinFET Xilinx RF-system on chip (RFSoC), which integrates eight ADCs, eight DACs, and considerable FPGA logic into a single chip; the entire GEP readout could potentially be reduced to 1 or 2 such chips. A board with this chip has been released\cite{ABACO2018}; initial power dissipation estimates are well below the GEP conservative assumption of 25 W per 1 GHz channel. Alternatively, mixed-signal application-specific integrated circuits (ASICs) that integrate ADCs and signal processing have been developed for similar applications\cite{hsiao20152}.


\section{GEP Surveys, Simulations, and Science}

\subsection{Surveys}

\begin{figure}
\begin{center}
\begin{tabular}{c}
\includegraphics[height=6.0cm]{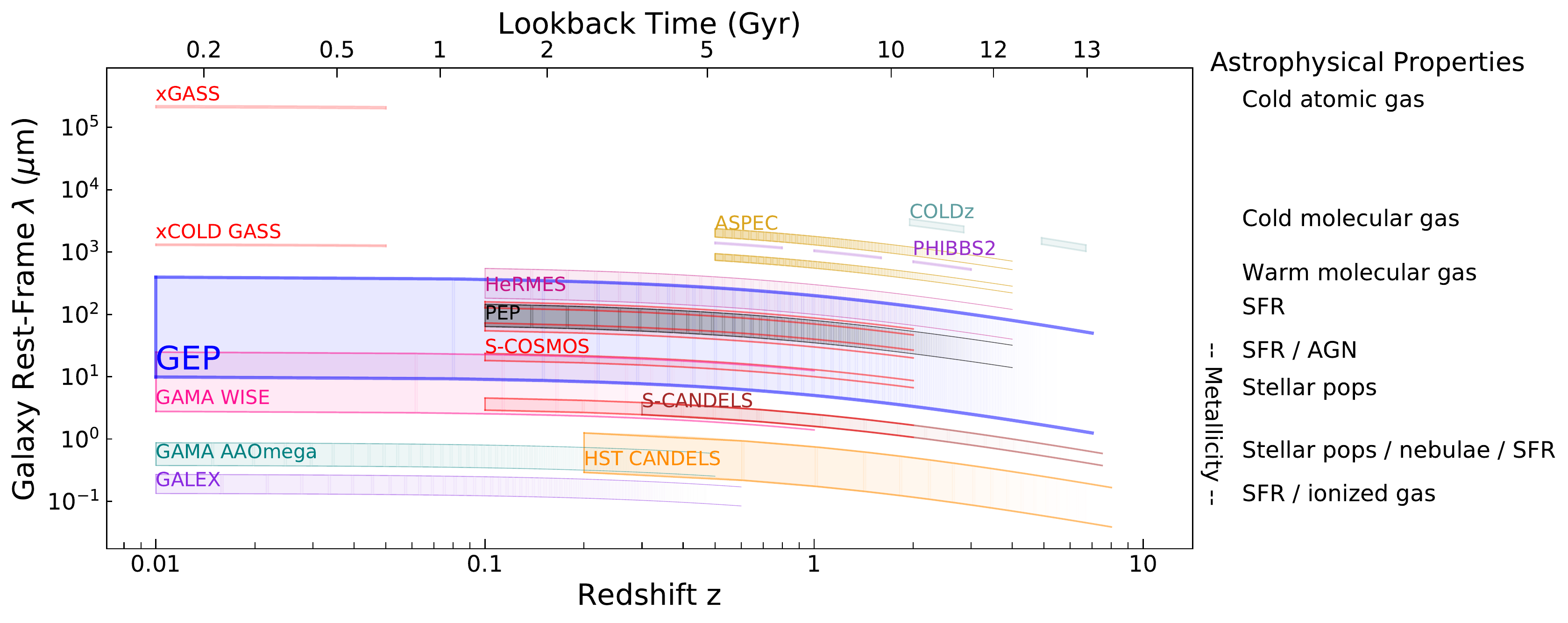}
\end{tabular}
\end{center}
\caption 
{ \label{fig:GalaxySurveyGEP}
Context for galaxy evolution surveys: a representative sample of galaxy surveys as a function of wavelength and redshift\cite{Koekemoer_2011, Ashby_2015, Sanders_2007, Bianchi_2017, Driver_2009, Cluver_2014, Mitchell_Wynne_2012, Catinella_2018, Saintonge_2017, Freundlich_2019, Pavesi_2018, 2016ApJ...833...67W}.  The emphasis is on space-based surveys and not all surveys are shown to minimize confusion in the figure.  Some ground-based surveys are included that provide critical, unique information. Completed or nearly completed surveys have solid borders, while planned surveys have dashed borders. The left terminations indicate either lower redshift limits arising from observational spectral bands or are indicative of the areas surveyed. For example, WISE covered the entire sky and therefore probed very low redshifts comprehensively, whereas deep, narrow surveys such as {\it HST} CANDELS and {\it Spitzer} CANDELS probed fewer galaxies at the lowest redshifts.  Terminations that fade to the right indicate reduced sensitivity to detecting galaxies at high redshifts because of flux limits.
} 
\end{figure}

\begin{figure}
\begin{center}
\begin{tabular}{c}
\includegraphics[height=6.0cm]{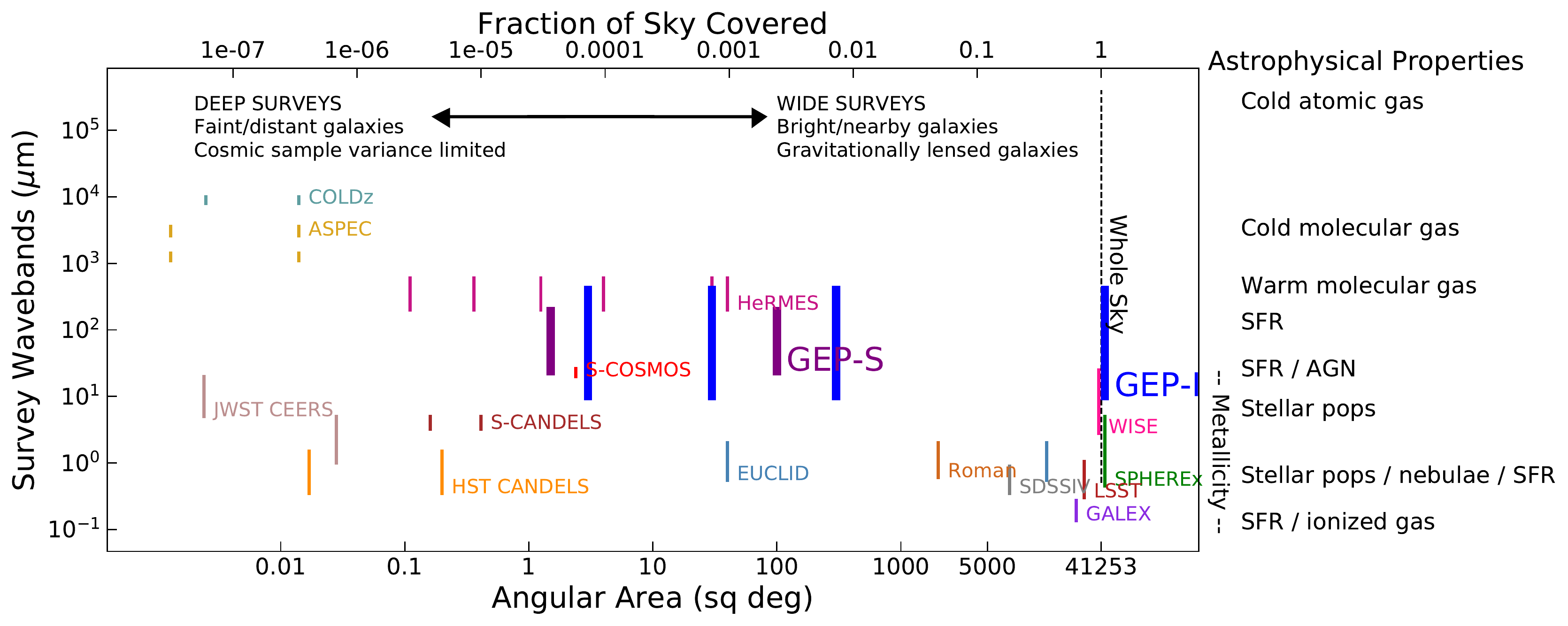}
\end{tabular}
\end{center}
\caption 
{ \label{fig:GalaxySurveyGEP_area}
Context for galaxy evolution surveys: a representative sample of galaxy surveys as function of wavelength and sky area\cite{Pavesi_2018, 2016ApJ...833...67W, Mitchell_Wynne_2012, Sanders_2007, Ashby_2015, Koekemoer_2011, Cluver_2014, Bianchi_2017, JWST-CEERS, Euclid, Roman, 2016AJ....151...44D, 2018AAS...23135421B, 2009arXiv0912.0201L}. Surveys on the left side of the panel are generally deep, optimized for detecting faint and high-redshift galaxies but are cosmic sample variance limited, whereas wide surveys toward the right generally detect brighter and lower-redshift galaxies (the exception is lensed galaxies, which are rare). GEP-I and GEP-S surveys indicated with the bold purple and blue lines occupy a broad range of wavelength and sky area space. {\it{IRAS}} covered the entire sky, but its observations were shallow (by comparison to GEP) and did not include spectra.  GEP surveys extend sensitive mid- to far-infrared imaging and spectroscopy more than two orders of magnitude from previous surveys, finally exploring the dusty side of galaxy evolution across the whole sky and over 2/3 the age of the Universe.
} 
\end{figure}

GEP is designed to do large, unbiased surveys. Its overarching goal is to provide a legacy dataset with broad utility for studying the evolution of star-formation, interstellar medium, and black hole growth in galaxies. 
Parameters of GEP's surveys in the context of other surveys are show in Figs. \ref{fig:GalaxySurveyGEP} and \ref{fig:GalaxySurveyGEP_area} with complementary surveys and the astrophysics that they probe.

\begin{figure}
\begin{center}
\begin{tabular}{c}
\includegraphics[height=5.5cm]{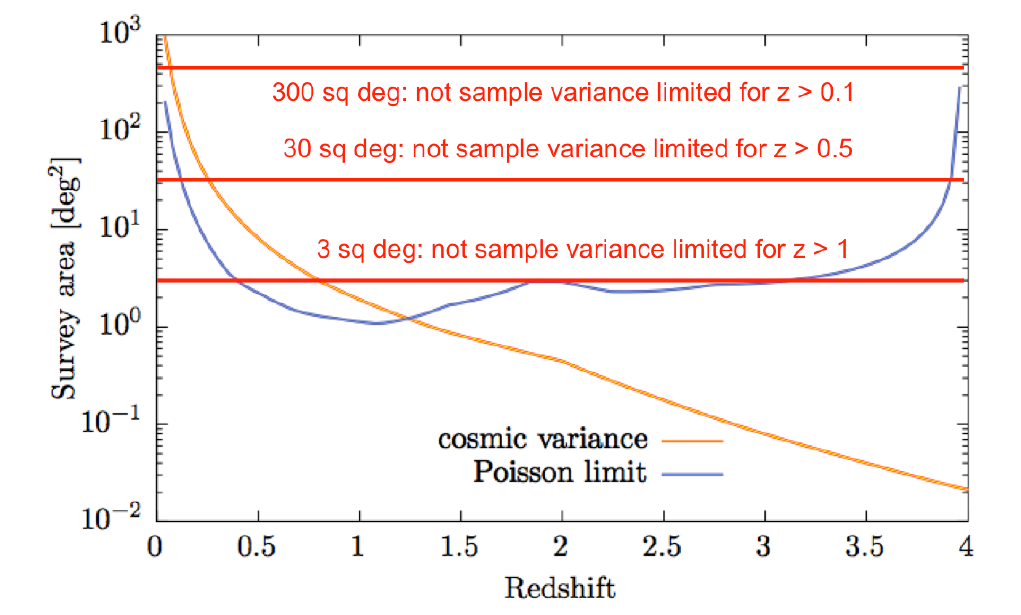}
\end{tabular}
\end{center}
\caption 
{ \label{fig:CosmicVariance}
Cosmic sample variance and the Poisson limit for GEP surveys. The red horizontal lines denote the depths of three of the four survey layers (excluding the all-sky survey). GEP's 'wedding-cake' survey layers observe sufficiently large angular areas that they will not be limited by cosmic sample variance in the target redshift ranges of each tier. The 3 and 30 square degree surveys will also not be limited by shot noise for $0.5 \le z \leq 3.5$.  The 300 square degree has a lower shot noise limit ($z \sim 0.1$) and the all-sky survey will extend to even lower redshifts.  The blue line shows the minimum areas needed to detect enough galaxies to enable luminosity function slope measurements to a precision of 10\% at the faint end.
} 
\end{figure}

The GEP-I survey program is optimized to sample a comprehensive range of redshifts and galaxy luminosities  (Table \ref{tab:SurveyYields}). A combination of four depths and areas will sample low redshift and rare, luminous galaxies, faint, high-redshift galaxies, and intermediate redshift and luminosity galaxies. Large numbers of galaxies will be detected in each tier, preventing sample variance and small number statistics from hampering conclusions on galaxy evolution with redshift (Fig. \ref{fig:CosmicVariance}). 
It is anticipated that the GEP-I surveys (except the all-sky survey) will be centered on and divided between the north and south ecliptic poles to minimize the photon backgrounds from primarily zodiacal dust and secondarily Galactic dust. This will provide
sky coverage overlap with Euclid and Nancy Grace Roman Space Telescope (NGRST) surveys,
which will provide near infrared photometry, morphologies, and stellar masses for galaxies detected by GEP.

\begin{figure}
\begin{center}
\begin{tabular}{c}
\includegraphics[height=5.5cm]{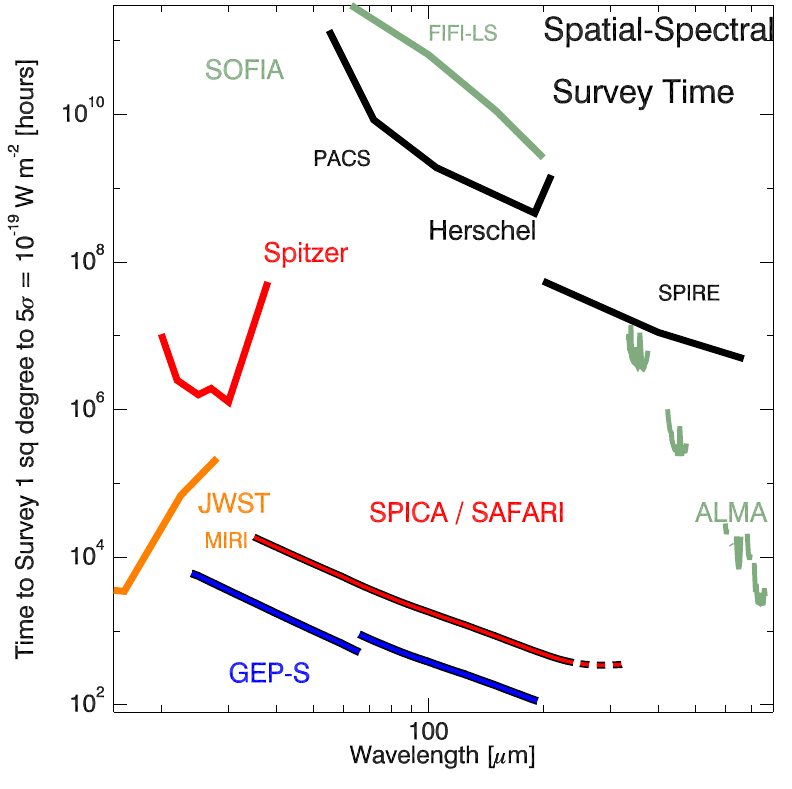}
\end{tabular}
\end{center}
\caption 
{ \label{fig:SpectroscopyDepths}
Spectral survey time to a given depth in the mid- and far-IR ({\it lower is faster}). The GEP-S spectrometer modules offer gains of 6 orders of magnitude in observing speed relative to state-of-the-art -- {\it Herschel} and SOFIA -- because of their warm apertures. The GEP-S speed exceeds that of SPICA because of the larger focal plane array format that enables long slits.  
} 
\end{figure}

Three types of spectroscopic measurements with GEP-S will complement GEP-I photometric surveys: (1) Individual observations of specific galaxies identified in the GEP-I surveys to provide precise redshifts and to validate PAH redshifts and to obtain measurements of the  far-infrared emission lines. (2) Unbiased spectroscopic surveys obtained by rastering GEP-S on the sky. 3) Spectral maps of nearby galaxies. GEP-S will perform a deep spectroscopic survey over 1.5 square degrees and a wide spectroscopic survey over 100 square degrees (Table \ref{tab:SurveyYields}). 
The spectral survey datasets will detect galaxies the far-infrared fine-structure transitions (and the continuum, when binned). The wide survey will be used to stack GEP spectra on the NGRST and/or Euclid grism sources to provide high signal-to-noise ratio average galaxy spectra in bins (Table \ref{tab:Stacking}). The GEP-S spectral mapping speed is shown in Fig. \ref{fig:SpectroscopyDepths}, where it is seen that GEP would be a substantial improvement of the state of the art with times to survey sky regions in the far-infrared six orders of magnitude faster than previous observatories.

\begin{table}[ht]
\caption{GEP galaxy surveys and yields obtainable with background-limited sensitivities.} 
\label{tab:SurveyYields}
\footnotesize{
\begin{center}       
\begin{tabular}{|c|c|c|c|c|c|} 
\hline
\rule[-1ex]{0pt}{3.5ex} Module & Area & Depth & Region & No. Galaxies & No. Redshifts \\
\hline\hline
\rule[-1ex]{0pt}{3.5ex} GEP-I & 3 sq deg & 10 $\mu$Jy & Ecliptic Poles & $10^6$ above $\sigma_{conf}$ & $10^5$ \\
\hline 
\rule[-1ex]{0pt}{3.5ex} GEP-I & 30 sq deg & 30 $\mu$Jy & Ecliptic Poles & $10^6$ above $\sigma_{conf}$ & $2\times10^5$ \\
\hline 
\rule[-1ex]{0pt}{3.5ex} GEP-I & 300 sq deg & 100 $\mu$Jy & Ecliptic Poles & $10^7$ above $\sigma_{conf}$ & $5\times10^5$ \\
\hline 
\rule[-1ex]{0pt}{3.5ex} GEP-I & 300 All sky & 1 mJy & ... & $10^8$ above $\sigma_{conf}$ & $10^6$ \\
\hline 
\rule[-1ex]{0pt}{3.5ex} GEP-S & 1.5 sq deg & $7\times10^{-20}$ W m$^{-2}$  & Ecliptic Poles & $2\times10^4$ & $2\times10^4$ \\
\rule[-1ex]{0pt}{3.5ex} & & @100~$\mu$m  & & for $1<z<2$ & \\
\hline 
\rule[-1ex]{0pt}{3.5ex} GEP-S & Local galaxy & $1\times10^{-10}$ W m$^{-2}$ & Distributed & 400 & N/A \\
\rule[-1ex]{0pt}{3.5ex} & mapping & ster$^{-1}$~@122~$\mu$m & & & \\
\hline 
\rule[-1ex]{0pt}{3.5ex} GEP-S & 100 sq deg & $3.5\times10^{-19}$ W m$^{-2}$ & Overlap NGRST / & $4\times10^5$ &  $5\times10^4$ \\
\rule[-1ex]{0pt}{3.5ex} & & @100~$\mu$m & Euclid grism fields & & \\
\rule[-1ex]{0pt}{3.5ex} & & & Intensity mapping & ... & \\
\hline 
\rule[-1ex]{0pt}{3.5ex} GEP-S & GEP-I galaxies & $3\times10^{-20}$ W m$^{-2}$ & Distributed & 300 & 300 \\
\rule[-1ex]{0pt}{3.5ex} & $1.2<z<3$ & & & & \\
\hline 
\end{tabular}
\end{center}
}
\end{table}


\begin{table}[ht]
\caption{GEP Stacking of IR Galaxy Datasets obtainable with background-limited sensitivities.} 
\label{tab:Stacking}
\footnotesize{
\begin{center}       
\begin{tabular}{|c|c|c|c|c|c|c|c|c|} 
\hline
\rule[-1ex]{0pt}{3.5ex}  Survey & Redshift of H$\alpha$ & \shortstack{Area \\ (deg$^2$)} & \shortstack{Flux Depth \\ (erg/s/cm\textsuperscript{2})} & \shortstack{Density \\ (deg$^{-2}$)} & \shortstack{Median SFR \\ (M$_{\odot}$/yr}) & \shortstack{L\textsubscript{IR} \\ (L$_{\odot}$)} & N\textsubscript{Total} & \# Stacks \\
\hline\hline
\rule[-1ex]{0pt}{3.5ex}  Euclid & 0.9-1.8 & 15,000 & 2.4x10\textsuperscript{-16} & 4,000 & 33 & 1.9x10\textsuperscript{11} & 60 million & 260  \\
\hline
\rule[-1ex]{0pt}{3.5ex}  NGRST & 1-2 & 2,200 & 1.0x10\textsuperscript{-16} & 10,000 & 15 & 6.8x10\textsuperscript{10} & 22 million & 950  \\
\hline
\end{tabular}
\end{center}
}
\end{table} 

\subsection{Theoretical Framework and Simulations}

\label{sect:Simulations}

To quantify the science yield of the GEP-I survey program, a mock survey was constructed using a combination of the Millennium N-body simulation \cite{Springel2005} to provide the distribution of large-scale structure and the
Galacticus semi-analytic model \cite{Benson2012} to populate that simulation with galaxies based on physical models. For each dark matter halo in the simulation volume, the star-formation rate and black hole accretion rate of the galaxy was computed.
Bolometric infrared luminosities from star-formation were estimated as $L_\mathrm{IR} = 2.6 \times 10^{45}$ (SFR/$M_\odot$ yr$^{-1}$) ergs s$^{-1}$ \cite{Calzetti2013}, and those due to AGN activity as $\epsilon \dot{M}_\bullet \mathrm{c}^2$ (with $\epsilon$ being the radiative efficiency computed by Galacticus from the black hole spin and accretion rate, and $\dot{M}_\bullet$ being the black hole accretion rate). Infrared spectra with the corresponding AGN fraction were then assigned to galaxies using the models of Dale et al.\cite{Dale2014} and normalized to give the computed total infrared bolometric luminosity. 
In the Dale et al.\cite{Dale2014} models the distribution of dust heating intensities (and, therefore, temperatures) is parameterized in terms of $\alpha_\mathrm{SF}$, the exponent of the power-law distribution of heating intensities. Values of $\alpha_\mathrm{SF}$ for model galaxies were drawn from a Gaussian distribution with mean, $\bar{\alpha}_\mathrm{SF}=1.75$, and dispersion, $\sigma_{\alpha_\mathrm{SF}}=0.25$, with no redshift dependence. A range of different possible values of $\bar{\alpha}_\mathrm{SF}$ and $\sigma_{\alpha_\mathrm{SF}}$ were explored, with the above values selected as those which result in the closest match between our simulated and observed numbers counts of galaxies over the 24 and $350~\mu$m wavelength range.\cite{Bethermin2010,magnelli2013deepest} 
These spectra were then used to compute broadband luminosities for model galaxies in each GEP-I band. Finally, a light cone from this synthetic catalog was extracted, corresponding to a 4 square degree area, from $z = 0$ to $3$, and observed fluxes were determined in all GEP-I bands for each galaxy.

The galaxy number counts estimated in this way are lower limits because these (and other) galaxy evolution models generically underpredict the number of very high luminosity galaxies\cite{Benson2003} ($\sim 10^{13} \mathrm{L}_\odot$). Although we attempted to tune the Galacticus model number counts to match observations from {\it Spitzer} and {\it Herschel} by judiciously choosing from the Dale et al. \cite{Dale2014} spectral library, at high flux densities (i.e., $> 1$ mJy at 160~$\mu$m) the model counts are almost an order of magnitude too low compared to observations\cite{magnelli2013deepest}. The agreement is much better at 24 and 70 $\mu$m, coming close to matching observations\cite{Bethermin2010}. Some of the bright observed galaxies that are not accounted for in the models likely result from gravitational lensing, although lensing is unlikely to account for the majority of the discrepancy. We adopt our model predictions for the detection rates with the understanding that they are likely conservative.

Noise was added to the simulated maps corresponding to the expected map depths 
after they were convolved with GEP beam sizes.  A small region of the maps stepping through the bands in animated in Fig. \ref{vid:satellite}.

\begin{video}
\begin{center}
{\includegraphics[height=6cm]{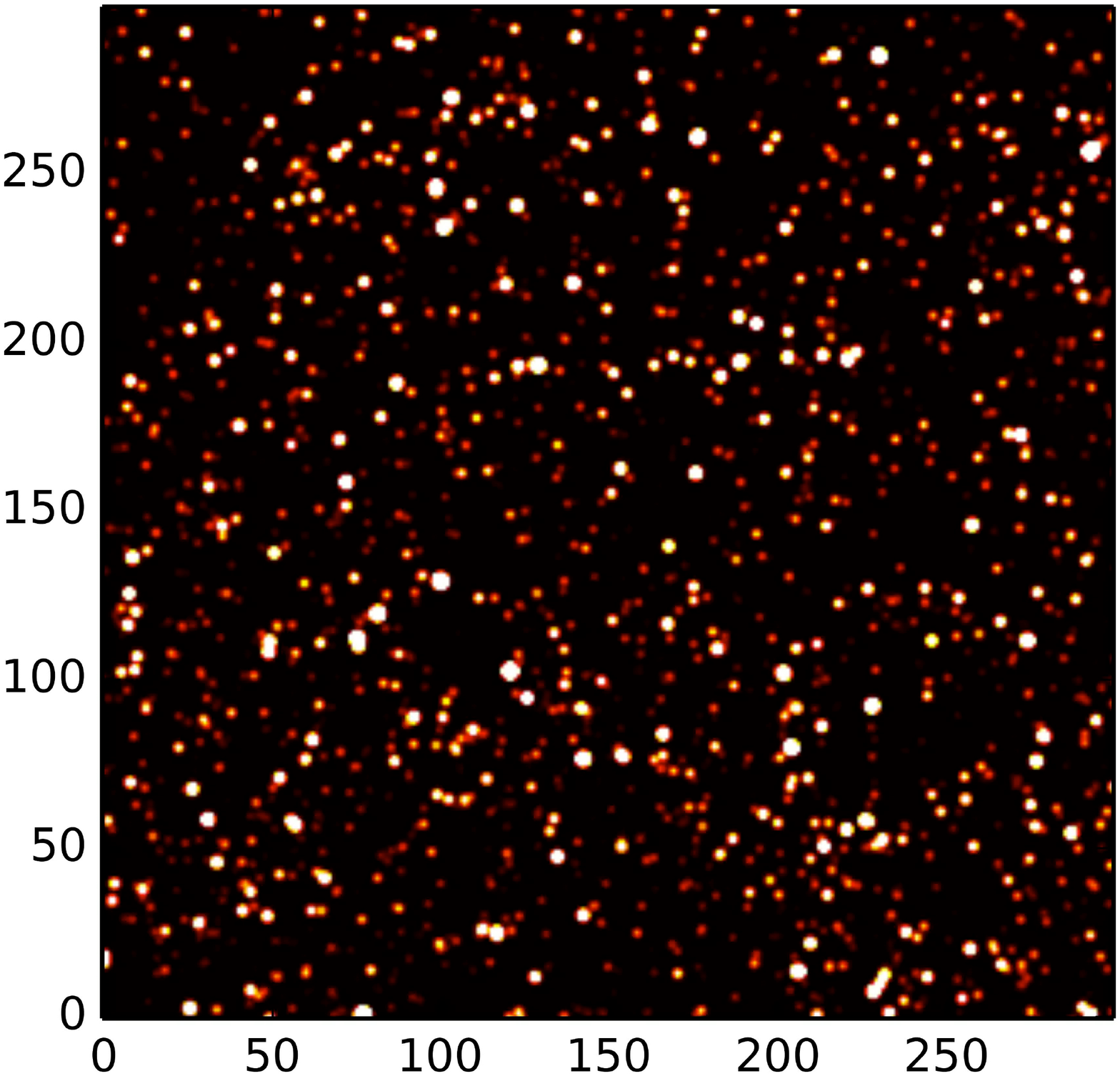}}
\\
\end{center}
\caption{\label{vid:satellite}A 10 $\mu$m still image from Video 1 (Video 1, MP4, 3.5 MB, which is in the on-line version), a rendering of the Galacticus simulations. The axis units are pixel number, with $3.43$\asec~$\times$ 3.43\asec~pixels -- the image is 17.2$'$  on a side. The animation runs continuously through the bands from 10 $\mu$m to 80 $\mu$m showing that galaxies alternately brighten and dim as PAH emission lines move through the bands and can thus be used to measure redshifts. }
\end{video}

\subsection{Extragalactic Source Confusion}


\label{sect:Confusion}

Extragalactic source confusion arises when images of galaxies -- the point spread function for unresolved galaxies -- overlap. It happens at far-infrared wavelengths where diffraction-limited beams can be several arcseconds or larger. 
Confusion 'noise' is the uncertainty in the extraction of a given source's flux due to the presence of a background of other sources which cannot accurately be subtracted from the signal.  Analytically, the confusion noise (1$\sigma$) is just the standard deviation of the flux from beam to beam, and for extraction of sources from a map, a `confusion limit' arises at 3--5$\times$ this 1$\sigma$ value.  This limiting flux below which extractions are unreliable, typically corresponds to that for which density of sources is 1 per several ($\sim$ 10) beams, depending on the source count relation.  This basic paradigm holds true when referring to extracting sources from a map without prior information, but there are well-developed methods to employ prior higher-resolution datasets to extract source parameters in a larger far-IR beam.  An excellent example is use of Spitzer IRAC and MIPS 24 micron positions to extract (statistically) fluxes in the 70 and 160 $\mu$m maps \cite{Papovich07spitzerstacking}. In this section we assess the important issue of source confusion with the GEP-I datasets, but first note that spectral surveys with GEP-S should not suffer appreciably from source confusion.

For spectral mapping with GEP-S, source confusion should not limit the extraction of line intensities. The key point is that the bright mid- and far-IR spectral lines are sparse and well-separated spectrally, and they form an unambiguous template for redshift identification.  Thus, each line can be conclusively identified and measured, even with multiple line emitting sources in the beam.  For a 3-D spatial-spectral survey, the analog of source counts for 2-D is line counts, the number of line emitters above a given flux level in a typical spatial - spectral bin of the survey.   This issue was studied for the Origins Space Telescope, which also features a wideband, moderate-resolving-power spectrograph; Bonato et al (2019)\cite{Bonato19OriginsSpectralSurveys} is a good reference.  As an example they predict that at 190~\mm (which is approximately GEP-S long-wavelength range limit), the density of lines with fluxes above 10$^{-20}\,\rm W\,m^{-2}$ (corresponding to a 25-hour integration with GEP-S) per beam-bin is 1/150 for Origins.  Correcting this for the larger GEP beam (2/5.9)$^{-2}$ and slightly smaller resolving power (200/300)$^{-1}$ results in an estimate of 1 source per 12 beam-bins for GEP-S at this wavelength.  This is approaching but not exceeding a practical confusion limit.  Shorter wavelengths of course are more forgiving with the smaller beam.

\begin{figure}
\begin{center}
\begin{tabular}{c}
\includegraphics[height=4.5cm]{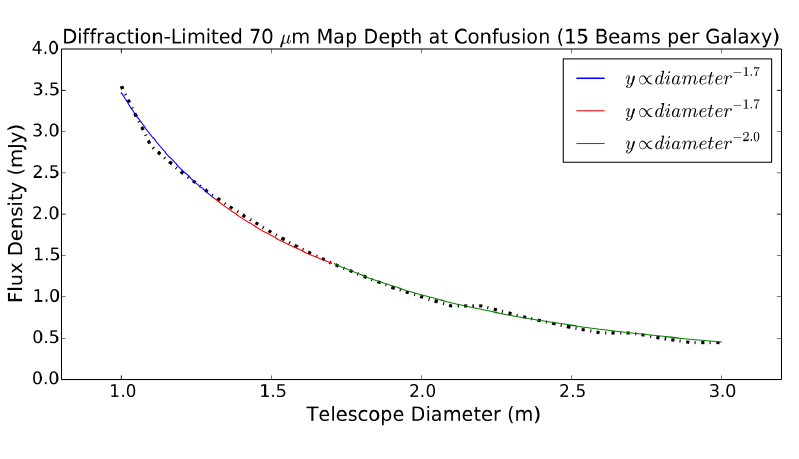}
\end{tabular}
\end{center}
\caption 
{ \label{fig:ApertureSize}
The flux density for 15 beams per galaxy as a function of telescope aperture diameter at $\lambda = 70$ $\mu$m. The number counts are from Béthermin et al.\cite{bethermin2012unified} and the curves are piecewise fits. GEP’s 2.0 m aperture is a 'sweet spot' between rapidly worsening confusion for smaller apertures and increasing cost for larger apertures.
} 
\end{figure}

We estimate the extragalactic confusion noise to be expected for GEP-I observations by considering confusion noise measurements from previous observations.   The confusion noise flux density was measured -- or upper limits were placed -- with {\it Spitzer} 70 and 160 $\mu$m\cite{MIPS2011, frayer2009spitzer} and {\it Herschel} 70, 100, and 250 $\mu$m\cite{magnelli2013deepest, berta2011building, Herschel2014}.  
We start by deriving an empirical relationship for confusion noise as a function of telescope aperture diameter to assess the dependence of confusion on telescope size and to check for consistency between {\it Spitzer} and {\it Herschel} observations, which had different wavebands and aperture sizes (0.85 m and 3.5 m, respectively).   Using the number counts models of Bethermin et al.\cite{bethermin2012unified}, the $\lambda = 70$ \micron~ flux density corresponding to 15 beams per source (where galaxies are not be substantially blended) scales approximately as the inverse of aperture diameter squared ($\propto D^{-2}$, slightly steeper for apertures smaller than 1.7 m, Fig. \ref{fig:ApertureSize}). 
Confusion noise measurements in the literature cited above scaled by this relationship are consistent, which validates the scaling relation for interpolation to the GEP 2.0 m aperture for estimation of confusion noise. Critically, the scaling also shows that aperture diameters below 2.0 m will be increasingly susceptible to confusion. Since the confusion flux density drops slowly with telescope diameter beyond 2.0 m, 2.0 m represents something of a `sweet spot' in the trade between confusion and cost.

A comparison of the expected GEP-I survey map depths for a 2.0 m aperture with the confusion measurements cited in the literature above yields the following:
 

\begin{enumerate}

\item At $\lambda = 24~\mu$m, GEP-I will not be confusion limited.

\item Astrophysically background-limited $\lambda = 70$ $\mu$m, $1\sigma$ root-mean-square (RMS) GEP-I map depths are 6 $\mu$Jy, 20 $\mu$Jy, 60 $\mu$Jy, and 6 mJy for the 3, 30, 300 square degree, and all-sky surveys, respectively. Scaling the {\it Spitzer} 300 $\mu$Jy, $\lambda = 70$ $\mu$m confusion noise by the empirical $D^{-2}$ relation yields 50 $\mu$Jy RMS. Thus, the noise in the deepest two surveys will be dominated by confusion noise at $\lambda = 70$ $\mu$m, the observational noise will just reach the confusion noise in the 300 square degree survey, and the all-sky survey will not be strongly affected by confusion noise.

\item At 100 $\mu$m, scaling by the $D^{-2}$ relation, {\it Herschel}’s observed confusion noise of 200 $\mu$Jy RMS would be 600 $\mu$Jy for GEP-I's 2.0 m aperture. Thus, the all-sky survey, with a $1\sigma$ map depth of 600 $\mu$Jy, would just reach the confusion noise level.

\item All four surveys will likely reach the confusion noise level at wavelengths longer than $\lambda = 100~\mu$m. However, observations should be made at these wavelengths to measure total luminosities of bright, low-z galaxies and lensed high-z galaxies.

\end{enumerate}

In short, GEP-I will likely have significant confusion noise at $\lambda = 70~\mu$m and longer, but not at shorter wavelengths. GEP-I will have to integrate deeper than the $70~\mu$m confusion noise in the two deepest surveys for PAH redshifts of $z 
\le 4$ galaxies with the wavebands at $50~\mu$m and below, which will not be limited by confusion noise. Additionally, monochromatic fluctuation ‘probability of deflection’ P(D) analyses show that it is possible to constrain galaxy populations meaningfully with observations deeper than the confusion noise level\cite{takeuchi2004, glenn2010hermes}. A polychromatic P(D) analysis with GEP-I observations covering the redshifted PAH features would be extremely powerful: it would yield precise galaxy number counts and redshifts statistically for the ensemble, and therefore luminosity functions, and tightly constrain galaxy evolution models. 
Furthermore, using cross-identification with counterparts at shorter wavelengths, galaxy properties can be measured even when there is source confusion. For example, Labbé et al.\cite{labbe2015ultradeep} showed that contamination by confusion can be reduced a factor of six with short-wavelength prior-based photometry, and positional and flux priors from ancillary data could enable improved spectral energy distribution extraction. Additionally, because the detector count in wavebands longer than 100 \micron~(GEP-I band 19) is small (\ref{sect:GEP-I}) compared to at shorter wavelengths, the cost of retaining them is merited for measurements of far-infrared luminosities, for nearby galaxy science that do not require observations as deep, and for P(D)-type fluctuation analyses.

\begin{figure}
\begin{center}
\begin{tabular}{c}
\includegraphics[height=4.5cm]{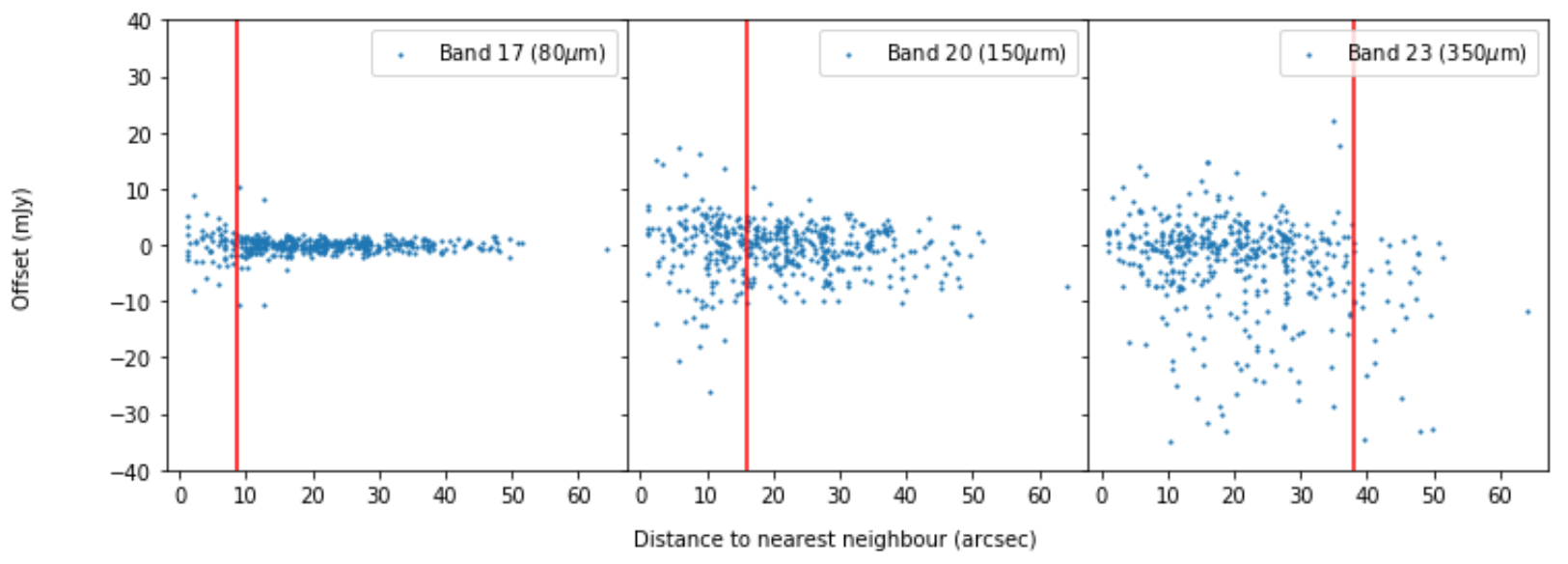}
\end{tabular}
\end{center}
\caption 
{ \label{fig:XID}
Differences between XID+ fitted and true galaxy flux densities as a function of distance to the nearest galaxy neighbor for three GEP-I long-wavelength bands: 17, 20, and 23 (with FWHM beam sizes of 8.6\asec, 16\asec, and 38\asec, respectively). Red vertical lines show the beam width as measured by the full width at half maximum. For bands 17 and 20, the fractional error in the flux density is small down to the beam size and clustered around zero. At the beam size and below the dispersion increases, although there is no bias. For band 23, GEP-I's longest-wavelength band, the simulation data do not extend to great enough separations to see that the offset decreases toward zero for separations above the beam size. In this band, there is a negative bias for some galaxies that will have to be corrected.
} 
\end{figure}

For wavelengths $\lambda \ge 70~\mu$m, corresponding to GEP-I bands 17 to 23, photometry of partly blended sources can still be extracted reliably. 
The Next Generation (X)Cross Identification (XID+) code was developed to estimate flux densities accurately from confusion-limited {\it Herschel} imaging\cite{Hurley2017}; we apply it here to GEP-I.  
The performance of XID+ was quantified by measuring the differences between fitted and true galaxy flux densities as a function of the distance to the nearest galaxy neighbor using the simulations described in Sect. \ref{sect:Simulations}. Positional priors were assumed from the mid-infrared wavebands and the fitted galaxy far-infrared flux densities were compared to their true values.  Fig. \ref{fig:XID} shows that down to galaxy separations of the beam size -- below the classical confusion ‘limit’ -- galaxy flux densities can be deblended with small fractional errors and little or no bias in most cases. Using the deblended sources, the map flux densities were reproduced with small residuals, although with poorer performance at the map edges because of missing galaxies outside the field of view. This will not be a concern for the majority of galaxies detected in GEP-I surveys, for which the minimum size is 3 square degrees.

\subsection{Anticipated Science}

\subsubsection{GEP-I Redshift Measurements}

To measure the cosmic star formation history of the Universe, GEP requires redshifts with precision $\sigma_z / (1 + z) \le 10$\% to $z = 2$, a requirement that our simulations show is achieved with margin. Generally, the redshift uncertainty is set by the width of the GEP-I bands relative to the widths of the PAH features. 
 
\begin{figure}
\begin{center}
\begin{tabular}{c}
\includegraphics[height=5.5cm]{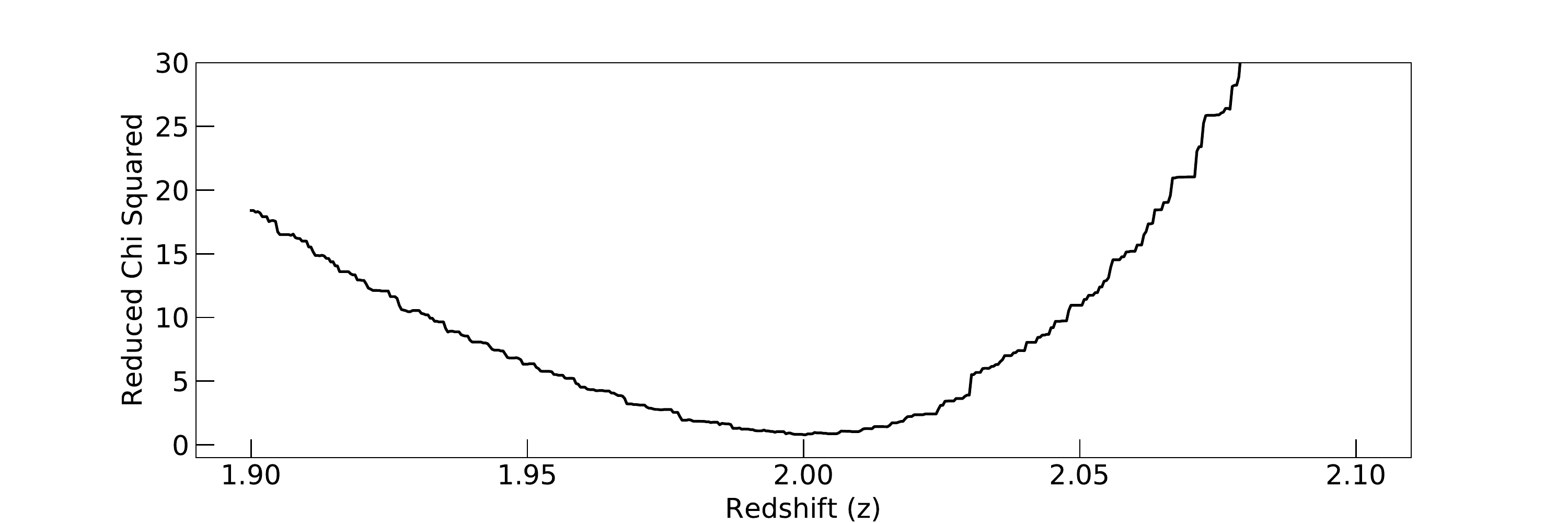}
\end{tabular}
\end{center}
\caption 
{ \label{fig:RedshiftChiSq}
The redshift reduced chi-squared distribution for a $z = 2$, L = $10^{12}~\mathrm{L}_{\odot}$ galaxy with the model spectrum shown in Fig. \ref{fig:GEP_Model_Spectra}.  The depth corresponds to the 30 square degree survey.   The measured redshift is $z = 2\pm0.02$, with $\sigma_z / (1 + z) = 0.01$.  This represents a best case because the observed spectrum was fit to the same model spectrum (same dust parameters) and ideal square $R = 8$ bandpasses were assumed.  Uncertainties from real GEP observations will be larger; however, this analysis shows that the multiple PAH emission lines combined with the deep silicate absorption can in principle lead to precise redshift determination.  The jagged nature of the chi-squared distribution arises from the sampling of the model spectrum.
} 
\end{figure}

Expected GEP-I redshift precision was estimated by adding noise to the model spectra according to each of the map depths given by the photon backgrounds. The mock spectra were binned into the GEP-I bands and $\chi_{\nu}^2$ were calculated by comparison to the spectral model and two other different spectral models: since the spectra of galaxies will not be known a priori, the comparison models were used to ascertain the uncertainties incurred by having a spectrum different from the model. 
Only the first eleven GEP-I bands ($10 - 37~\mu$m) were used for the redshift measurements 
because the steeply rising mid-infrared dust spectra influence the redshifts and the dust temperatures will not be known a priori.  A sample GEP-I redshift chi-squared distribution is shown in Fig. \ref{fig:RedshiftChiSq}. The nominal model had strong PAH emission features and the alternate comparison models had: (1) strong PAH emission features but cooler dust (hence a more slowly rising spectrum with wavelength) and (2) hot dust that substantially overwhelmed the PAH features above $10~\mu$m. The results were as follows:

\begin{enumerate}

\item Redshifts are obtainable for $10^{11}$ L$_{\odot}$ galaxies (corresponding to a median stellar mass of $4 \times 10^{9}\mathrm{M}_\odot$) out to $z = 2$ in the 3 square degree survey
with $\sigma_z / (1 + z) \le 0.1$.

\item Redshifts are obtainable for $10^{12}$ L$_{\odot}$ galaxies (corresponding to a median stellar mass of $1 \times 10^{10}\mathrm{M}_\odot$) to $z = 2$ in the 300 square degree survey and to $z = 4$ in the 30 square degree survey with $\sigma_z / (1 + z)\le 0.1$.

\item Redshifts are attainable for $10^{13}$ L$_{\odot}$ galaxies (corresponding to stellar masses in excess of $10^{11}\mathrm{M}_\odot$) to $z = 4$ in the 300 square degree survey 
$\sigma_z / (1 + z) \le 0.05$.

\item The redshift uncertainties are a function of redshift, galaxy luminosity, map depth, and strength of the PAH features relative to the continuum.

\item In the case of very warm dust, which represents an extreme case of entirely AGN-dominated galaxies, the redshifts have large uncertainties. However, even in this case, redshifts become detectable at $z > 2$ for deep surveys and luminous galaxies because the dust continuum is not strong below rest-frame $10~\mu$m and there is a strong 3 \micron~emission line, which is redshifted well into the GEP-I bands, and silicate absorption can still be present.

\end{enumerate}

A comprehensive set of simulations is required to quantify redshift measurements as a function of GEP-I spectral resolving power, AGN fraction and other sources of weak PAH emission (e.g., low metallicity), galaxy-to-galaxy variability, and instrument parameters such as non-ideal bandpass shapes and $1/f$ noise.  These simulations likely represent best-case scenarios.  These effects will be explored as a subsequent effort.

\subsubsection{GEP Redshift Measurements for Galaxies with Strong AGN}

GEP has two means for measuring redshifts.  One, GEP's $R = 200$ spectrometer GEP-S will provide redshifts for AGN-dominated galaxies via (low- and) high-ionization fine-structure lines.  Two, even in the instances of weak PAH-to-continuum emission in strongly AGN-dominated galaxies, the $10~\mu$m rest-frame silicate absorption feature will still enable redshift measurements.  
The harsh environments immediately around AGNs can in theory destroy PAHs,\cite{1991MNRAS.248..606R, 1992MNRAS.258..841V} but observationally, even galaxies with AGN often display these features, although they are weaker.   Detailed {\it Spitzer} IRS spectra of hundreds of galaxies with AGN activity found that some PAH features are seen in most cases,\cite{2010ApJ...709.1257T} showing that PAHs can survive in some of these environments.  Li\cite{2020NatAs...4..339L} summarizes the PAH observations in AGN by noting that the 6.2, 7.7 and 8.6 $\mu$m PAH features, although suppressed with respect to the 11.3 $\mu$m band (perhaps because smaller PAHs are more susceptible to the hard radiation field of an AGN) nevertheless can be present.  Li's Figure 5\cite{2020NatAs...4..339L} of the relative PAH line strengths shows the influence of radiation hardness, as well as the limits of current modeling in Seyfert galaxies, many of which show strong neutral PAH emission. 

From a theoretical perspective, coauthors of this paper have examined the role of AGN activity in a galaxy mid-IR SEDs in another paper\cite{McKinney2021} using a Gadget-2 simulation of an idealized (non-cosmological) major merger of two identical disk galaxies (Springel et al. 2005)\cite{2005MNRAS.364.1105S}.  The galaxies had initial halo and baryonic masses of $9\times10^{12}$ and $4\times10^{11}$ M$_{\odot}$, and a central black hole mass of $1.4\times10^5$ M$_{\odot}$. The star formation and feedback were modeled as described in Springel \& Hernquist (2003)\cite{2003MNRAS.339..289S} and Springel et al. (2005)\cite{2005MNRAS.364.1105S}. 
The radiative transfer code SUNRISE\cite{2006MNRAS.372....2J} was used to compute SEDs for seven viewing angles every 10 Myr throughout the simulation run, and by varying the efficiency of the AGN it was effectively turned off or raised to 100\% (for Eddington-limited Bondi-Hoyle accretion). The ratio of the two resultant SEDs reveals the influence of the pure AGN (see also Figure 9 in Dietrich et al. 2018\cite{2018MNRAS.480.3562D}).  Results are presented in McKinney et al. (2021)\cite{McKinney2021}, Figure 1 of which shows that the silicate absorption feature is present and will enable redshift measurements even when galaxies' luminosities are strongly dominated by AGN.


Other physical effects will also influence the strengths of these mid-IR PAHs features, for example low metallicity in high-z, young galaxies, or destructive supernovae activity.\cite{2020NatAs...4..339L} Although weak PAH features make photometric redshift measurements more challenging, it is precisely the complexity and variability of these features that 
ultimately 
enable diagnosis of the
physical conditions present. 


\subsubsection{Galaxy Luminosity Functions, Star Formation History, Separating AGN, and Clustering}
\label{sect:GalaxyDetections}

\begin{figure}
\begin{center}
\begin{tabular}{c}
\includegraphics[height=6.5cm]{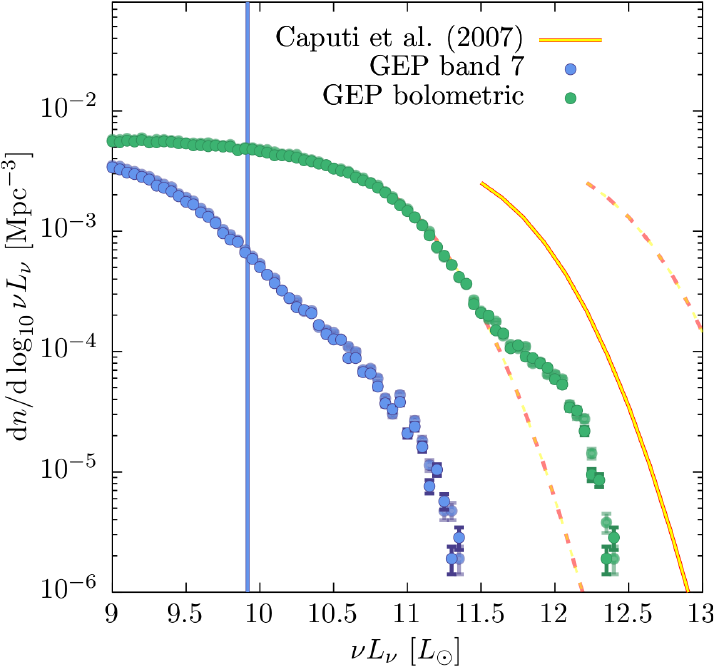}
\end{tabular}
\end{center}
\caption 
{ \label{fig:LumFuncs}
GEP luminosity function for $1.0 < z < 1.2$ derived from the Galacticus mock catalogs assuming a $\sigma_z / (1 + z) = 0.1$ uncertainty on galaxy redshifts. Error bars are estimated assuming Poisson statistics scaled from the 4 square degree area of our mocks to 30 square degrees. Blue points show the luminosity function in the GEP-I Band 7 (21.2 to 24.0 $\mu$m), while green points show the bolometric infrared luminosity function. The bolometric luminosity function is not shown below $10^9$ L$_{\odot}$ because it becomes incomplete due to the resolution of the simulation. The blue line is the $3\sigma$ detection limit. Faint points indicate the luminosity function that would be obtained if spectroscopic redshifts were available -- it is almost indistinguishable from that constructed using photometric redshifts. The orange lines are from Caputi et al.\cite{caputi2007}
} 
\end{figure}

GEP's science objectives require luminosity functions over a range of redshifts to observe the changes in galaxy formation and the build-up of stellar mass over cosmic time.  Faint-end (below L$^*$) mid- and far-infrared luminosity function slopes have not been measured above $z = 0.5$ and there is disagreement about the faint-end slopes even at $z \sim 0$.  GEP-I’s surveys will detect hundreds of millions of galaxies and measure redshifts for millions of them for derivation of luminosity functions.
With GEP-I surveys, faint-end slopes below Log$_{10}$ (L$_{IR}$/L$_{\odot})$ $= 11$ for $z = 0.5$, below Log$_{10}$ (L$_{IR}$/L$_{\odot}$) $= 11.5$ at $z = 1$, and below Log$_{10}$ (L$_{IR}$/L$_{\odot})$ $= 12$ at $z = 2$, will be measured. Fig. \ref{fig:LumFuncs} shows current observational determinations of the infrared bolometric luminosity function and compares these to a sample luminosity function derived from our mock GEP catalogs. 


\begin{figure}
\begin{center}
\begin{tabular}{c}
\includegraphics[height=6.5cm]{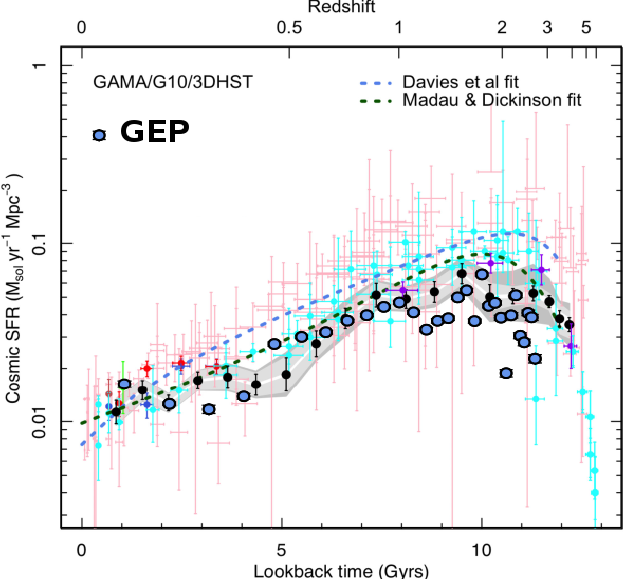}
\end{tabular}
\end{center}
\caption 
{ \label{fig:SFH}
A recent study of the cosmic star formation density history from deep multiband data -- for which the principal source of uncertainty is the limitation of the far-infrared data -- reproduced from Fig. 13 of Reference \cite{driver2018gama} with GEP simulation data overplotted. Blue circles show expectations from a combination of GEP full sky, 300, 30, and 3 square degree surveys (for $0 < z < 3$ only), showing that GEP will measure the star formation history with unprecedented precision. In most cases error bars are smaller than the symbols.  The large scatter in the GEP points arises from simulations of limited cosmic volume.  While the error bar sizes are appropriate for the GEP surveys, the scatter in the GEP survey data should be very small compared to what is seen here.
} 
\end{figure}

One of GEP's primary goals is to quantify the star formation history of the Universe by utilizing very large statistical samples of galaxies that overcome Poisson statistics and sample variance from large-scale structure.  How well will it do?  Fig. \ref{fig:CosmicVariance} showed that Poisson statistics and sample variance will be overcome and Fig. \ref{fig:LumFuncs} demonstrated that luminosity functions will be well quantified even below L$^*$ at $z = 1$.  How do these translate into measurement of the history of cosmic star formation rate density?  Fig. \ref{fig:SFH} shows that the star formation history will be quantified with uncertainties an order of magnitude below the state of the art to at least $z = 3$.  Uncertainties will likely improve substantially beyond $z = 3$ also; however, that was not included in the simulations (section \ref{sect:Simulations}).
Because it maps large areas, GEP will also be effective at detecting gravitationally lensed galaxies.  This will enable it to serendipitously detect sub-L* galaxies at $z = 2$ with a tail out to higher redshifts, perhaps $10^4$ lensed galaxies in the all-sky survey and 50 in the deeper 300 square degree survey.

\begin{figure}
\begin{center}
\begin{tabular}{c}
\includegraphics[height=4.0cm]{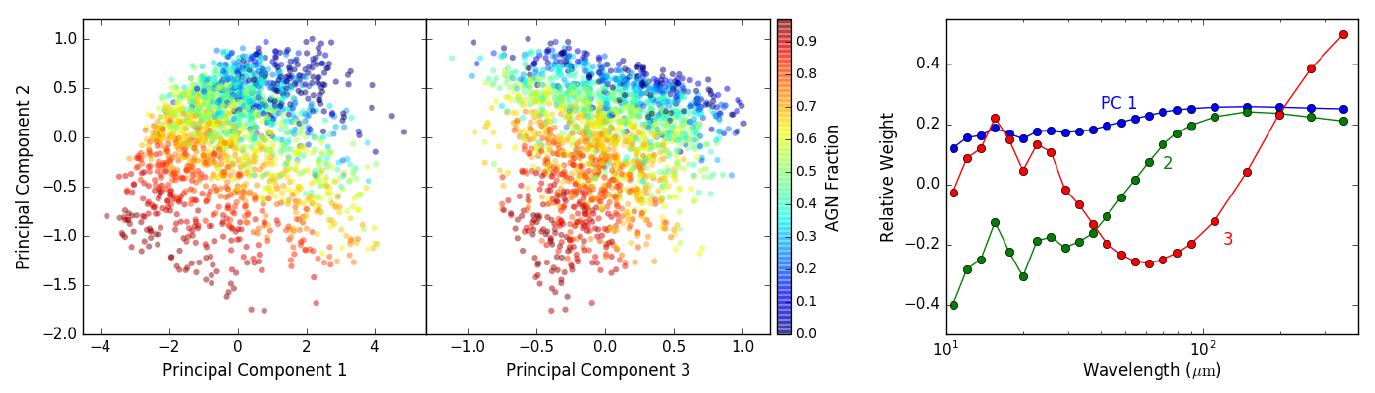}
\end{tabular}
\end{center}
\caption 
{ \label{fig:PCA}
GEP-I's multiband observations will enable the relative contributions of galaxy infrared luminosities from AGN and star formation to be discriminated based on their different mid- and far-infrared spectra. Left: Outcome for analysis of the 3 square degree survey at redshift $z = 1.0\pm0.1$ when all GEP bands are detected. Results are similar for other survey depths and subsets of GEP bands provided adequate detection rates. Right: The three principal components identified by the algorithm. Together, they measure the AGN fraction, radiation field hardness, and the redshift.
} 
\end{figure}

Another GEP goal is to identify embedded AGN and quantify their contributions to the infrared luminosities of galaxies.  The mid-infrared part of the spectrum provides excellent leverage for this because the dust is hotter -- thus the spectra are 'bluer' -- and the PAHs are more subsumed by the bright dust continuum.  Fig. \ref{fig:PCA} shows that a simple principal component analysis with only three components is sufficient to identify AGN and separate their contributions to infrared luminosities from star formation with GEP-I data.  The high-ionization lines observed with GEP-S will indicate the presence of AGN and enable the larger catalog of GEP-I galaxies to be calibrated against the fine-structure line luminosities.  These observations can be used to assess AGN influence on galactic interstellar media and the coevolution of supermassive black holes and star formation.

\begin{figure}
\begin{center}
\begin{tabular}{c}
\includegraphics[height=6.0cm]{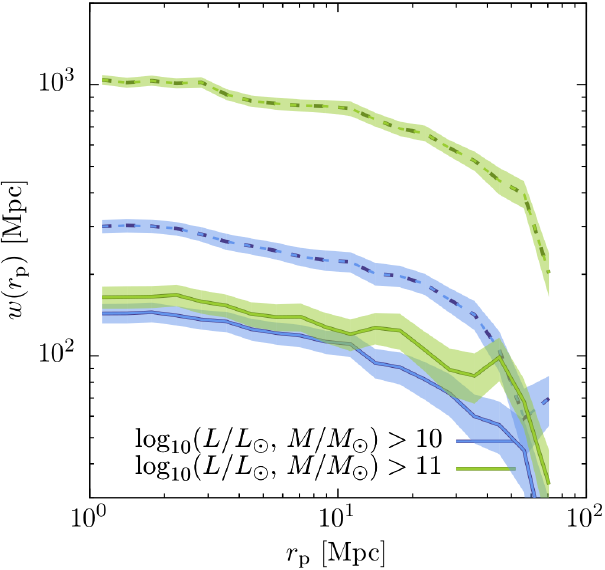}
\end{tabular}
\end{center}
\caption 
{ \label{fig:TwoPointCorr}
Projected correlation functions from the mock catalog for galaxies selected by bolometric infrared luminosity (solid lines) and stellar mass (dashed lines) in a $1.0 < z < 1.2$ redshift slice. Bands indicate the expected $1\sigma$ uncertainties for the 30 square degree survey. The Galacticus simulations show what is already known: galaxy mass correlates strongly with halo mass, leading to a strong dependence of clustering strength with mass. Conversely however, it predicts that bolometric infrared luminosity (which indicates star-formation rate) is weakly correlated with halo mass because even galaxies in low mass halos can have occasional strong starbursts, leading to a weak dependence of clustering strength on infrared luminosity. 
} 
\end{figure}

Within a $\Lambda$CDM cosmology, the clustering of galaxies reveals the masses of dark matter haloes that they occupy.  Galaxy mass correlates strongly with halo mass; however, it is possible that mid- and far-infrared luminosities of galaxies correlate more weakly with halo mass because even low-mass galaxies can be temporarily infrared-bright from star formation bursts.  Clustering of GEP-I catalogs will strongly test galaxy formation and evolution theories that predict halo occupations as a function of redshift (Fig. \ref{fig:TwoPointCorr}).

\subsubsection{The Milky Way and Nearby Galaxies}

\begin{figure}
\begin{center}
\begin{tabular}{c}
\includegraphics[height=10.5cm]{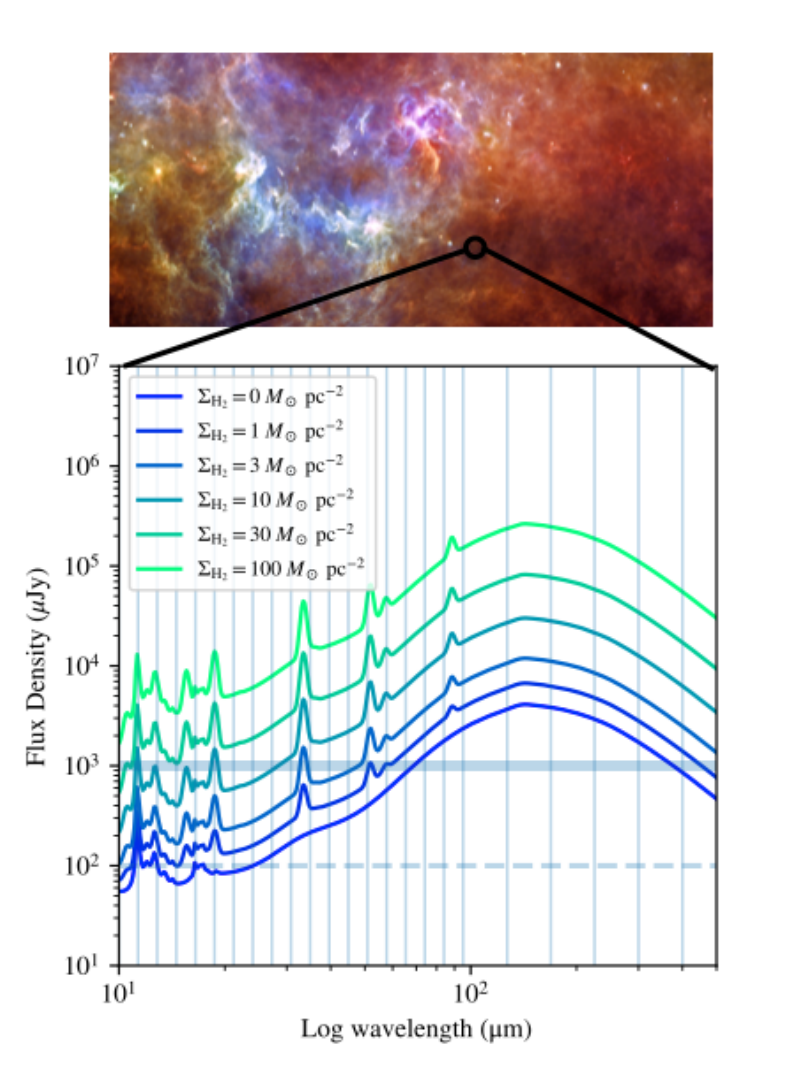}
\end{tabular}
\end{center}
\caption 
{ \label{fig:MilkyWay}
GEP-I will provide spatially resolved spectral energy distribution mapping across the star-forming interstellar medium over $10^{7.5}$ lines of sight in the Milky Way and nearby galaxies, providing a large data set to understand how local environment establishes interstellar conditions. The figure shows how the dust emission from the Aquila molecular cloud (top) would be mapped into spatially resolved spectral energy distributions (bottom) over the GEP-I bands (band edges denoted with blue vertical lines) with flux density estimates in 3 arcsecond apertures. The curves correspond to the expected signatures for varying levels of molecular gas in a stellar population with a stellar surface density of $300$ M$_{\odot}$ pc$^{-2}$.  The spectral slopes, fine-structure line-to-continuum ratios, and PAH feature strengths vary depending on the gas surface densities.  The horizontal lines indicate the approximate $5\sigma$ depths of the all-sky survey (solid) and 300 sq deg survey (dashed). The depth flux densities are lower limits because they do not account for the increased background in the Galactic Plane.    Additionally, detector nonlinearities from large backgrounds have not been modeled or accounted. 
} 
\end{figure}

GEP will map the Milky Way and the Magellanic Clouds with GEP-I during the all-sky survey.  The sensitivity required of the telescope and instruments for the deep extragalactic observations will enable high signal-to-noise ratio mid- and far-infrared spectral energy distributions of star-formation regions.  The 3.43\asec~ mid-infrared resolution will resolve gradients in dust temperatures and PAH excitation (Figure \ref{fig:MilkyWay}), thereby probing the radiation fields and chemistry, while the far-infrared spectral energy distributions will yield luminosities of embedded stars, dust temperatures, and gas masses.

\subsubsection{Interstellar Physical Conditions}

Like the multiband imager, GEP-S is designed for rapid surveys but targeting the gas-phase spectral features in the rest-frame mid- and far-infrared.  The spectra from GEP-S will chart the cosmic history of the interaction of galaxies' energy sources (stars and supermassive black holes) with their gas reservoirs.  Transitions from low-ionization states such as [C II], [N II], [O III], and [NeII], can be used to infer the masses of interstellar neutral and ionized gas components, the hardness of the ultraviolet radiation fields and its implications for the stellar initial mass function, and the density of HII regions from which the pressure in the ISM can be inferred. 
Intermediate ionization transitions of nitrogen and oxygen, [NIII] and [OIII], indicated the degree of stellar processing, a proxy for metallicity \cite{nagao2009next} in galaxies irrespective of the dust and gas temperature uncertainties.
Finally, spectral tracers of highly ionized gas, such as [S IV] at 10.5 $\mu$m, [Ne V] at 14.3 and 24.3 $\mu$m, and [O IV] at 25.9 $\mu$m, with ionization potentials of 35–97~eV, when combined with bright lines with much lower ionization potential (e.g., [Ne II] at 12.7 $\mu$m), indicate the relative amount of heating from young stars and AGN.

To utilize this basic spectroscopic toolkit, there are three types of  observations envisioned for GEP-S: 1) ‘Blind’ field-filling spectroscopic surveys obtained by rastering GEP-S on the sky, described further below; 2) individual pointed observations of specific galaxies identified in the photometric surveys or other facilities, for both precise redshifts to validate the photometric technique and fluxes in the full suite of far-infrared spectral features; and 3) spectral maps of nearby galaxies to study feedback and interstellar energy balance in detail.

Survey spectroscopy with GEP-S is a particular strength of the GEP long-slit spectrometer architecture, and the study envisions two spectral surveys: a deep survey over 1.5 square degrees and a wide spectroscopic survey over 100 square degrees.  The resulting datasets will be used in at least three ways. First, they will detect $10^4$ galaxies directly in the far-infrared fine-structure transitions (and the continuum when binned).  Second, the wide survey will be used to stack on the Nancy Grace Roman Space Telescope (NGRST) and/or Euclid grism sources to provide high signal-to-noise ratio average galaxy spectra in bins. Both surveys will be excellent datasets to  measure the total cosmic luminosity density in the various far-infrared lines and ratios among integrated line intensities.

The synergy with the coming near-infrared grism surveys will be an especially powerful tool.  Comparing the anticipated Euclid and NGRST grism depths\cite{merson2018predicting} with GEP depth in the 100-square-degree survey suggests that stacks will access populations down to 19 (7) $\times\,10^{10}\,\rm L_{\odot}$, respectively, in a total sample of 60 (22) million galaxies, respectively for Euclid and NGRST.  The sensitivity is sufficient to form hundreds of independent bins in quantities such as star formation rate, stellar mass, and metallicity. The resulting spectra will measure not just the bright fine-structure lines, but the `second-tier' transitions such as [Ne V] (at 14 and 24 $\mu$m rest frame), which provide dust-immune AGN indicators, and the quadrupole pure rotational H$_2$ lines (28, 17, 12, ...\mm) which indicate warm shocked gas \cite{Ogle14ShockedH2}.  In considering the Euclid and Roman datasets, those sources which are very highly obscured may not yield clear redshifts with the GRISM --- they of course cannot be included in the stacking analysis. These sources will, however, make excellent candidates for pointed spectroscopy with GEP-S, which will access the bright and dust immune far-IR fine-structure transitions of [OI] and [OIII].  At a redshift of $z = 1$, the 1-hour GEP-S sensitivity of 6$\times$10$^{-20}\,\rm W\,m^{-2}$ is a good match to a $\sim$10$^{11}\,\rm L_{\odot}$ source whose less-obscured counterpart is included in the stacking analysis.  While a sample of $\sim$100 such sources will not have the full statistical power of the stacking analyses, it will nevertheless provide useful and otherwise inaccessible constraints on the properties of these obscured galaxies.

These spectral surveys will also be used for line intensity mapping, a powerful emerging technique in which the clustering of line-emitting galaxies is detected as fluctuations in a 3-D spatial-spectral dataset with redshift encoded as wavelength.  The key virtue is that technique is sensitive to all forms of emission, not just the individually detected galaxies, so it offers the potential for absolute aggregate measurements irrespective of a single-source detection limit. GEP-S intensity mapping of mid- and far-infrared fine structure lines -- and possibly PAHs -- will yield signals in autocorrelation, but it will also be powerful when cross-correlated with rest-frame optical and ultraviolet surveys and ground-based millimeter-wave surveys that separately probe stars and the atomic and molecular interstellar medium. 

\subsubsection{Probing Feedback from Extraplanar Gas}

A hallmark feature of the cryogenic telescope and sensitive GEP-S spectrometer is excellent surface brightness sensitivity.  This will provide a breakthrough capability for assessing feedback effects in local galaxies.  Feedback in the form of stellar and AGN winds and supernova explosions are believed to eject gas and dust from star-forming galaxy disks, creating a reservoir of low-column-density gas in the outskirts of galaxies.  This material exists below the threshold for star formation (hydrogen column density below 1-10$\,\rm M_{\odot}\,cm^{-2}$) and has been difficult to measure.  GEP-S will be capable of detecting [CII] and [NII] in this material, even when the local density is well below the critical density for excitation.  Sensitive maps of the diffuse gas can be obtained with GEP-S around low-redshift star-forming galaxies and AGN. These maps will provide an powerful test of the theories and models of stellar feedback which have evolved to have great detail but need testing by observations.

\subsection{Future Science Exploration}


GEP's science reach will be far broader than outlined here. For example, monochromatic probability-of-density -- P(D) -- fluctuation analyses have shown to be effective for constraining galaxy number counts into the confusion noise\cite{takeuchi2004, glenn2010hermes}.  Multicolor P(D) analyses could go much further:  distinguishing between galaxy number count models and thereby constraining luminosity functions as a function of redshift to substantially greater depth than for analyses restricted to galaxies whose brightness are above individual detection thresholds.  In nearby galaxies, mid- and far-infrared spectral energy distributions are needed for panchromatic spectral energy distribution fitting to constrain stellar populations, nebular conditions, star formation rates, and identify embedded AGN on a galaxy-by-galaxy basis.  In the Milky Way, mapping of PAH emission in various environments, such as the vicinities of hot, young stars and photodissociation regions will probe the radiation fields and PAH production, excitation, and destruction.  
Lastly, with the possibility of linear-variable filters with $R > 8$, the optimum resolution for PAH redshifts should be revisited. Then, GEP-I redshift precision should be quantified as a function of galaxy luminosity, redshift, AGN fraction (with strong AGN mid-infrared continuum leading to weak PAH lines), and instrumental parameters, such as spectral resolving power $R$.

Finally, it must be noted that the GEP design reference mission is designed for dedicated surveys.  The surveys should be designed with community input and consideration of the rich landscape of imminent multi-wavelength galaxy surveys.  And, while the design reference mission does not support a open-time phase to manage costs, with no expendable cryogens the expected lifetime of the mission exceeds the planned survey durations.  An extended mission with guest-observer opportunities would yield a large volume of science, as the {\it Spitzer} extended mission has shown.

\section{Conclusions}

The Galaxy Evolution Probe concept was developed to make precise measurements of the star-formation rates, nuclear accretion rates, and interstellar conditions of galaxies over cosmic time and over the full range of cosmic environments.  It was designed as a survey mission to obtain large, well-defined samples of galaxies that will not be limited by counting statistics nor by cosmic sample variance.  Its science goals will be realized with 2.0 m, cryogenic telescope ($\le 6$ K) and two instrument modules, each with arrays of approximately 25,000 kinetic inductance detectors.  It will utilize established cryogenic, telescope, and bus technologies.  GEP will improve the measurement uncertainties of the cosmic star formation rate density by more than an order of magnitude out to a redshift of at least $z = 3$, the current extent of our predictive simulations.  Obscured AGN in galactic centers will be identified on a galaxy-by-galaxy basis, yielding new luminosity functions.  Critical measurements of metallicities and interstellar conditions will be obtained for galaxies, complementary to probes at other wavelengths.  GEP's ability to provide these powerful new observational capabilities derives from emerging mid- and far-infrared detector technologies that require continued development and nurturing to realize not only GEP, but a future flagship opportunity such as OST. 


\subsection*{Disclosures}
The authors have no relevant financial interests in the manuscript and no other potential conflicts of interest to disclose.

\subsection* {Acknowledgments}
This work was supported, in part, by a NASA Astrophysics Probe Concept Study grant to Jason Glenn (NASA award number NNX17AJ89G). This research was funded in part by the Jet Propulsion Laboratory, California Institute of Technology, under a contract with NASA. Jason Glenn thanks JPL and Ball Aerospace for their strong support of the Study effort. The GEP team would like to thank many contributors whose important work helped lead to the successful completion of this study but whose roles traditionally do not qualify for inclusion in authorship lists of scientific publications: engineering aides, documentarians, administrators, administrative aides, and reviewers.


\bibliography{report}   
\bibliographystyle{spiejour}   


\vspace{2ex}\noindent\textbf{Jason Glenn} Since 2020, Jason Glenn has been a Research Astrophysicist, Stellar, Galactic, and Extragalactic, in the Observational Cosmology Laboratory at the NASA Goddard Space Flight Research Center.  Prior to that, he was a professor of astrophysics at the University of Colorado Boulder for 20 years. He received his BS degree in physics from the University of New Mexico in 1991 and his PhD degree in astronomy from the University of Arizona in 1997.  He is the author of more than 200 journal papers. His current research interests include galaxy evolution, the cool interstellar medium in galaxies, and infrared through millimeter-wave instrument and detector development. 

\begin{figure}[!h]
\begin{tabular}{c}
\includegraphics[height=7.5cm]{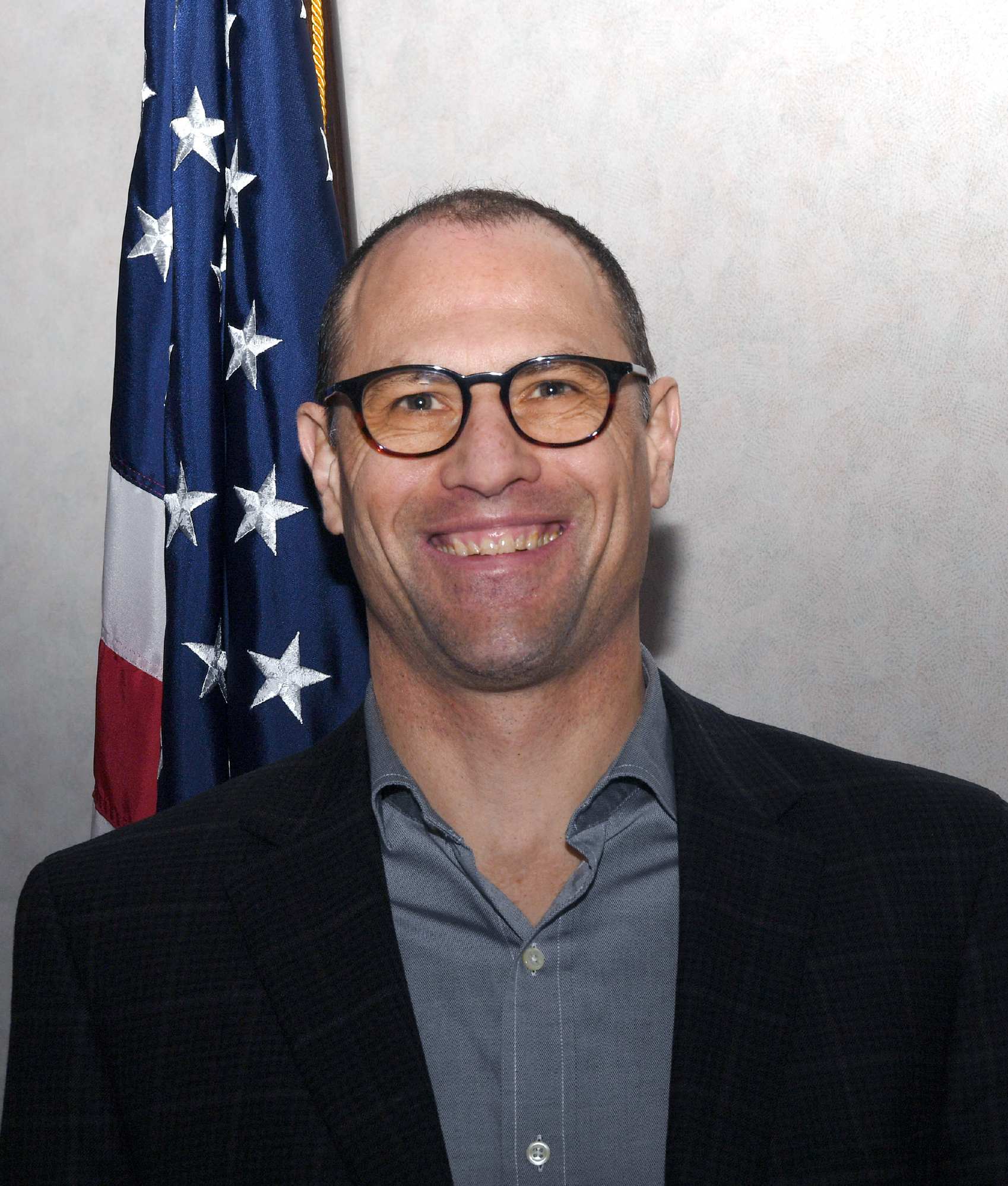}
\end{tabular}
\end{figure} 

\vspace{2ex}\noindent\textbf{C. Matt Bradford}	Charles Matt Bradford received his doctorate from Cornell in 2001. He held a Millikan postdoctoral fellowship at Caltech from 2001 to 2003 and has been on the science staff at JPL since that time. He enjoys developing and fielding new submillimeter- and millimeter-wave instrumentation, and using spectroscopy to study interstellar medium conditions in galaxies. Current projects include a balloon-borne far-IR spectrometer, and an on-chip mm-wave spectrometer, and ultra-sensitive far-IR detectors for cryogenic space missions.

\vspace{2ex}\noindent\textbf{Rashied Amini} Rashied Amini, Ph.D., is a systems engineer at Jet Propulsion Laboratory, California Institute of Technology, working in mission formulation and in research of autonomous technologies. Recently, he was the HabEx and Galaxy Evolution Probe Study Lead, submitted to the 2020 Astrophysics Decadal Survey. As a result of his formulation work, he is interested in supporting the maturation of technologies critical to science exploration. He received a Ph.D. in Physics from Washington University in St. Louis.

\vspace{2ex}\noindent\textbf{Lee Armus}  Lee Armus is a senior staff scientist at IPAC with over 25 years of experience in infrared spectroscopy and imaging. He is currently the lead scientist at the Roman Space Telescope Science Support Center, and formerly the lead for the IRS Instrument Support Team at the Spitzer Science Center. His research is focused on several topics central to the science of the Galaxy Evolution Probe, including galaxy mergers, galactic outflows, and luminous infrared galaxies.
h.	Raphael Shirley is a postdoctoral researcher at the University of Southampton. 

\vspace{2ex}\noindent\textbf{Andrew Benson}  Andrew Benson is a Staff Scientist at the Observatories of the Carnegie Institution for Science. His research focuses on understanding the nature of dark matter and the process of galaxy formation, with a particular emphasis on formulating a coherent picture of the many different aspects of these problems. Benson has developed a model of dark matter and galaxy formation physics, Galacticus, which blends both analytic understanding and numerical techniques.

\vspace{2ex}\noindent\textbf{Jeremy Darling} Jeremy Darling is a professor at the University of Colorado.  He studies cosmology, galaxy evolution, massive black holes, and astrophysical approaches to fundamental physics.  

\vspace{2ex}\noindent\textbf{Jeanette Domber}  Jeanette L. Domber is a program manager for Ball Aerospace Civil Space. She holds a PhD in Aerospace Engineering from the University of Colorado at Boulder.

\vspace{2ex}\noindent\textbf{Sarah J. Lipscy} Sarah J. Lipscy is the Deputy Director of Civil Space Business Development at Ball Aerospace. She holds a PhD in Astronomy and Physics from UCLA.

\vspace{2ex}\noindent\textbf{Raphael Shirley} Raphael Shirley is a postdoctoral researcher at the University of Southampton. His research is focused on galaxy evolution and the role of active galactic nuclei in star formation over cosmic time. He has worked on Herschel imaging, multi-wavelength astronomy, and supernova observations. He is now working on image processing pipelines for the Vera C. Rubin Observatory and the VISTA telescope. He has previously held positions at Sussex, Cambridge, and the IAC in Tenerife.

\vspace{2ex}\noindent\textbf{Howard A. Smith} Howard A. Smith is a Senior Astrophysicist at the Center for Astrophysics, Harvard \& Smithsonian (CfA) in Cambridge, with over 350 published scientific articles, and is a member of the Harvard Astronomy Department.  His research field is the origins of stars in the Milky Way and other galaxies, with specialization in techniques of infrared spectroscopy and instrument development.

\vspace{2ex}\noindent\textbf{Jonas Zmuidzinas}	Jonas Zmuidzinas (B.S. Caltech 1981, Ph.D. Berkeley 1987) joined the Caltech physics faculty in 1989 and currently serves as the Merle Kingsley Professor of Physics and Director of the Caltech Optical Observatories. He has also concurrently held positions at JPL as a Senior Research Scientist (2006-16), Director of the JPL Microdevices Laboratory (2007-11), and JPL Chief Technologist (2011-16). His research focuses on superconducting detectors and devices and their application to astronomy.

\vspace{1ex}
\noindent Biographies and photographs of the other authors are not available.

\listoffigures
\listoftables

\end{spacing}
\end{document}